\documentclass[a4paper,11pt]{article}

\usepackage{jheppub} 

\usepackage[T1]{fontenc} 

\usepackage{epsf,epsfig,subfigure,graphicx,amsmath,amssymb}

\usepackage{amsfonts}
\usepackage{latexsym}
\usepackage{color}

\newcommand{\dis}[1]{\begin{equation}\begin{split}#1\end{split}\end{equation}}
\newcommand{\be}{\begin{equation}}
\newcommand{\ee}{\end{equation}}
\def\bea{\begin{eqnarray}}
\def\eea{\end{eqnarray}}

\newcommand{\eq}[1]{Eq.~(\ref{#1})}

\newcommand{\VEV}[1]{\langle #1 \rangle}

\newcommand\tev{\,{\rm TeV}}
\newcommand\gev{\,{\rm GeV}}

\newcommand\ev{\,{\rm eV}}

\title{Phenomenology in supersymmetric neutrinophilic Higgs model with sneutrino dark matter}

\author[a,1]{Ki-Young Choi\note{Corresponding author.}}
\author[b]{Osamu Seto}
\author[c]{Chang Sub Shin}

 \affiliation[a]{Korea Astronomy and Space Science Institute, Daejon 305-348,  Republic of Korea}
 \affiliation[b]{Department of Life Science and Technology, Hokkai-Gakuen University, Sapporo 062-8605, Japan}
 \affiliation[c]{New High Energy Theory Center, Department of Physics and Astronomy, Rutgers University, Piscataway NJ 08854, USA}

\emailAdd{kiyoungchoi@kasi.re.kr}
\emailAdd{seto@physics.umn.edu}
\emailAdd{changsub@physics.rutgers.edu}

\abstract{
We study a supersymmetric neutrinophilic Higgs model
 with large neutrino Yukawa couplings
 where neutrinos are Dirac particles and the lightest right-handed (RH) sneutrino is the lightest supersymmetric particle (LSP)  as a dark matter candidate.
Neutrinophilic Higgs bosons need to be rather heavy by
 the precise determination of the muon decay width and dark radiation constraints
 for large Yukawa couplings.
From the Large Hadron Collider constraints, neutrinophilic Higgsino mass need to be heavier
 than several hundred GeV or close to the RH sneutrino LSP mass.
The latter case is interesting because
 the muon anomalous magnetic dipole moment
 can be explained
 with a relatively large lightest neutrino mass, 
 if RH sneutrino mixings are appropriately fine tuned
 in order to avoid
 stringent lepton flavor violation constraints.
Dark matter is explained by asymmetric RH sneutrino dark matter
in the favoured region by the muon anomalous magnetic dipole moment.
In other regions, RH sneutrino could be an usual WIMP dark matter.
}

\begin{document}

\maketitle
\flushbottom

%

\section{Introduction}
\label{introduction}

A scalar field could be responsible for the breakdown of a large gauge symmetry
and the generation of the masses of gauge bosons and fermions.
In fact, a scalar boson discovered at the Large Hadron Collider (LHC) appears to be 
 the Higgs boson in the standard model (SM) of particle physics~\cite{Aad:2012tfa,Chatrchyan:2012ufa}.
After the spontaneous symmetry breaking, the mass of each fermion, namely quarks and charged leptons,
 is given by the product of each Yukawa coupling constant
 and the vacuum expectation value (VEV) of the Higgs field.

However,
 one might suspect a special mechanism for the generation of neutrino masses
 and a special reason for its smallness, because
 masses of neutrinos are very small compared with other SM fermions.
One approach is the so-called seesaw mechanism with very heavy right-handed (RH)
 Majorana neutrinos, where the smallness of neutrino mass can be understood
 as a consequence of the high scale of RH neutrino mass~\cite{Seesaw}.
 
Another approach is the  neutrinophilic Higgs model~\cite{Ma,Wang,Nandi}.
In this model,
 neutrino has Dirac mass terms generated by another Higgs field whose
 VEV is much smaller than that of the SM Higgs,
 where the smallness of neutrino mass is a consequence of the smallness of
 the other Higgs VEV.
In this case, 
 the neutrino Yukawa couplings
 can be much larger than those with only the SM Higgs field if the Higgs VEV is small,
 because the neutrino mass is the product of the VEV of the neutrinophilic Higgs field and the Yukawa couplings. 
 
The large neutrino Yukawa couplings in the
 neutrinophilic Higgs model shows many interesting features, such as
 the possibility of the RH sneutrino as thermal dark matter (DM)~\cite{Choi:2012ap,Choi:2013fva,Mitropoulos:2013fla}
 or the low scale thermal leptogenesis~\cite{HabaSeto,HSY}. 
In addition, the large Yukawa couplings have more implications in the flavour structures and astrophysical phenomenon. 

In this paper, we examine various phenomenological aspects of the large Yukawa interactions
 in the supersymmetric extended neutrinophilic Higgs model. 
Those include the anomalous magnetic moments of muon, lepton flavour violation, 
 experimental constraints on the couplings and the masses of new particles,
 and cosmological and astrophysical constraints
 including indirect detection signatures by asymmetric sneutrino DM
 through gamma ray and neutrinos.

In Section~\ref{CandI:Pheno},
 we consider the constraints on the Yukawa couplings from neutrino masses and mixings,
 the muon decay, collider searches, and lepton flavour violation. 
Taking these constraints into account,
 we study the possibility to explain the muon anomalous magnetic moment in this model.
In Section~\ref{Constraints:Cosmo}, we consider the cosmological constraints from dark radiation, and
 in Section~\ref{Sec:darkmatter} we study the possibility of the lightest RH sneutrino as DM and 
 the astrophysical constraints on the models. We conclude our study in Section~\ref{sec:con}.
We provide some formulas in the Appendices.

\section{Phenomenological constraints and implications}
\label{CandI:Pheno}

In a supersymmetric model, the interaction is described by the term in superpotential
\begin{equation}
W_N= (\nu_R^c )_i (y_{\nu})_{i\alpha} L_\alpha\cdot H_\nu \,
\label{WYukawa}
\end{equation}
where $L_\alpha$ is the lepton doublet of the Standard Model, $\nu_{Ri}$ is a gauge singlet RH neutrino superfield
 and $H_\nu$ is a scalar doublet in addition to the standard two Higgs doublets in the MSSM and $i,\alpha=1,2,3$ denotes the generation index.
In the so called neutrinophilic Higgs model with the neutrino Dirac mass
 given by a small VEV of neutrinophilic Higgs field, $\VEV{H_\nu^0}$,
 neutrino Yukawa couplings can be as large as of the order of unity.  

We consider the Yukawa interaction of Dirac neutrino and neutrinophilic Higgs as
\dis{
{\mathcal L} = \bar{\nu}_{Ri} (y_\nu)_{i\alpha} L_\alpha H_\nu+  \textrm{h.c.}\, \label{Yukawa}
}
After the electroweak symmetry breaking,
 the neutrinophilic Higgs field develops the VEV $\VEV{H^0_\nu}=v_\nu/\sqrt{2}$
 and generates the neutrino mass.  
Since neutrinos are Dirac particles in this model,
 their mass matrix is simply proportional to $3 \times 3$ neutrino Yukawa coupling matrix.
The neutrino mass matrix, or equivalently Yukawa interaction, is given by
\begin{eqnarray}
&&{\cal L}_{\nu mass} = \bar{\nu}_{R i}  (m_{\nu})_{i \alpha} \nu_{L\alpha} +\textrm{h.c} , 
\end{eqnarray}
with 
\begin{eqnarray}
&& (m_{\nu})_{i \alpha} = (y_{\nu})_{i \alpha} \frac{ v_{\nu} }{\sqrt{2}} ,
\end{eqnarray}
here and hereafter we assume the Yukawa couplings are real, for simplicity.
Therefore, the left- and right- handed neutrinos compose the four component Dirac mass eigenstates.

In a supersymmetric theory, there exists the Yukawa interaction of scalar RH neutrino with
 the same Yukawa coupling of \eq{Yukawa}, given by
\dis{
\mathcal{L} = \tilde{\nu}_{Ri}  ( y_{\nu})_{\, i\alpha} \bar{L}_\alpha P_R \tilde{H}_\nu + \textrm{h.c.}.
\label{Yukawa_scalar}
}
Here, RH sneutrinos $\tilde{\nu}_{R i}$ are defined
 as the superpartner of each $\nu_{R i}$.
If the lightest RH sneutrino is the lightest supersymmetric particle (LSP),
it can be a good candidate for DM as shown in Refs~\cite{Choi:2012ap,Choi:2013fva}. 
Throughout this paper, we consider this case of RH sneutrino DM
 and study the phenomenological constraints and implications.

\subsection{Neutrino mass and mixing}
\label{neut.yukawa}

Without loss of the generality, 
we can regard that the $ \nu_{R i} $ is already mass eigenstate.
The neutrino mass matrix is diagonalized with 
the Maki-Nakagawa-Sakata (MNS) matrix $U^{\rm{ MNS}}$, which transfers
 LH neutrinos from mass eigenstates ($\nu_{L,i}$) to flavor eigenstates ($\nu_{L,\alpha}$) $\nu_{L,\alpha}=(U^{\rm{ MNS}})_{\alpha i}\nu_{L,i}$,
\begin{equation}
 {\rm diag}(m_1, m_2,m_3)_{ij} =  (m_{\nu})_{i \alpha}(U^{\rm{ MNS}})_{\alpha j}.
\end{equation}

The neutrino oscillation data gives two independent mass squared differences and three mixing angles~\cite{PDG},
\dis{
& m_2^2-m_1^2 \simeq 7.5 \times 10^{-5} {\rm eV}^2,  \\
& |m_3^2-m_1^2| \simeq 2.3 \times 10^{-3} {\rm eV}^2, \\
& \sin^2 2\theta_{23} >  0.95,  \\
& \sin^2 2\theta_{12} \simeq 0.857,  \\
& \sin^2 2\theta_{13} \simeq  0.095. 
}
The neutrino Yukawa couplings can be expressed in terms of these neutrino oscillation parameters as

\begin{eqnarray}
 (y_{\nu})_{i \alpha}
 &=&  \frac{\sqrt{2}}{ v_{\nu} } {\rm diag}(m_1, m_2,m_3)_{ij}(U^{\rm{ MNS}})_{j \alpha }^{-1} \nonumber \\
 &\simeq & \frac{1}{ v_{\nu} }\left( \begin{array}{ccc}
\sqrt{2} m_1\cos\theta_{12}  & -m_1 \sin\theta_{12} & m_1 \sin\theta_{12} \\
\sqrt{2} m_2\sin\theta_{12} & m_2 \cos\theta_{12} & -m_2 \cos\theta_{12} \\
\sqrt{2} m_3\sin\theta_{13} & m_3 & m_3
\end{array} \right) .
\end{eqnarray}
Here, we neglect for simplicity any CP phase and take $y_{\nu}$ to be real.
In the second line, $\theta_{23}\simeq \pi/4$ and $\sin\theta_{13} \ll 1$ are used, 
 while we use the full formula of MNS matrix in our numerical calculation.

The upper bound on the sum of neutrino masses is provided
 by cosmological arguments.
Since its value strongly depends on data set and a cosmological model used in the analysis,
 for reference, we here just quote one of less conservative values
\begin{equation}
 \sum m_{\nu} <0.23\, {\rm eV},
\end{equation}
 from Ref.~\cite{Ade:2013zuv}.

In Table~\ref{tab:benchmarks}, we list four benchmark points used in our analysis
 for a given lightest neutrino mass,$m_1=0,\, 0.07\ev$ for normal hierarchy ($m_1 < m_2<m_3$)
 and $m_3=0,\, 0.05\ev$ for inverted hierarchy ($m_3 < m_1<m_2$) of neutrino masses.
To estimate Yukawa coupling constants, we have to fix one extra free parameter $v_{\nu}$.
Example value sets of $y_{\nu}$ in Tab.~\ref{tab:benchmarks}
 are obtained under the assumption that the largest coupling is unity. 
The actual value can be somewhat larger or smaller by changing $v_{\nu}$.

\begin{table}[!t]
\begin{center}
\begin{tabular}{|c|c|c|c|c|}
\hline
{\bf Point}&{\bf 1: normal }&{\bf 2: normal}&{\bf 3: inverted}&{\bf 4: inverted} 
\\
\hline
\hline
$m_{1,3}$ [eV] &$m_1= 0.0$ &$m_1= 0.07$ & $m_3=0.05$ &$ m_3=0.0$\\
\hline
\hline
$(y_{\nu})_{i \alpha}$
& 
$
\left({\small
\begin{array}{ccc}
0.0 & 0.0 & 0.0 \\
0.14 & 0.13 & -0.16 \\
0.22 & 1.0 & 1.0 
\end{array}} 
\right)
$
& $\left( \small{
\begin{array}{ccc}
0.96 & -0.57 & 0.36 \\
0.65 & 0.62 & -0.77 \\
0.22 & 1.0 & 1.0
\end{array}}
 \right)$
& $\left( \small{
\begin{array}{ccc}
1.0 & -0.59 & 0.37 \\
0.68 & 0.65 & -0.80 \\
0.14 & 0.62 & 0.62
\end{array}}
 \right)$
& $\left( 
\begin{array}{ccc}
1.0 & -0.59 & 0.37 \\
0.68 & 0.65 & -0.80 \\
0.0 & 0.0 & 0.0
\end{array}
 \right)$ \\
\hline
\hline
$\sum m_{\nu}$ [eV] & 0.06 & 0.23 & 0.19 & 0.10 \\
\hline
\hline
$v_{\nu}$ [eV] & 0.05 & 0.08 & 0.08 & 0.05 \\
\hline
\end{tabular}
\end{center}
 \caption{
Neutrino Yukawa coupling matrix of the four benchmark 1, 2, 3 and 4 used in this work
 are shown. 
The resultant $\sum m_{\nu}$ for a given input parameter $m_1$ or $m_3$ is also listed.
$v_{\nu}$ is a free parameter. The noted values of $v_{\nu}$ are examples
 in the case that we normalize the largest element of Yukawa matrix to be $1.0$.}
\label{tab:benchmarks}
\end{table}

\subsection{Muon decay width}
\label{muondecay}

Muon decays into electron, electron-type anti-neutrino and muon-type 
 neutrino, $\mu \rightarrow e \bar{\nu}_e \nu_{\mu}$ in the Standard Model.
However, in the neutrinophilic Higgs model, the Yukawa interaction \eq{WYukawa} gives additional contribution to the SM prediction
  via one loop induced vertices
and changes the muon decay width.
Those come from box diagrams with a loop 
 ($\nu_R, H_{\nu}^0$ and $H_{\nu}^-$) or 
 ($\tilde{\nu}_R, \tilde{H}_{\nu}^0 $ and $\tilde{H}_{\nu}^-$).
Moreover,
 muon has additional decay mode into electron, RH neutrino and anti-RH neutrino,
 through the $H_{\nu}^+$ mediation.
This new contribution can be severely constrained by the precise measurement of the decay rate and inverse decay rate of muon.

Now we are going to estimate those additional contributions to muon decay width.
For this, we define mass eigenstates $\tilde{N}_i$ from flavour states $\nu_{Ri}$ with an unitary matrix $S_{ij}$ as defined by
\begin{eqnarray}
 \tilde{\nu}_{R i} &=& S_{ij} \tilde{N}_j, \label{def:sneu:masseigenstate} \\
  S_{ij} &=& R(\sigma_{12})R(\sigma_{23})R(\sigma_{13}), \label{def:S} 
\end{eqnarray}
 with $R$ being a rotation matrix and the variables, $\sigma_{12}, \sigma_{23}, \sigma_{13}$, are the corresponding mixing angle.
The Yukawa interaction \eq{Yukawa_scalar} becomes
\dis{
\mathcal{L} = \tilde{N}_{k}(S^T)_{ki}  ( y_{\nu})_{\, i\alpha} \bar{L}_\alpha P_R \tilde{H}_\nu + \textrm{h.c.}, 
\label{Yukawa_scalar_mass}
}
 for the mass eigenstates. 
Thus, we define $\tilde{y}_{i\alpha}\equiv (S^T y_{\nu})_{i\alpha}$ for Yukawa couplings of the RH sneutrino, lepton and Higgsino.

\begin{figure}[!t]
  \begin{center}
  \begin{tabular}{cc}    
  \includegraphics[width=0.4\textwidth]{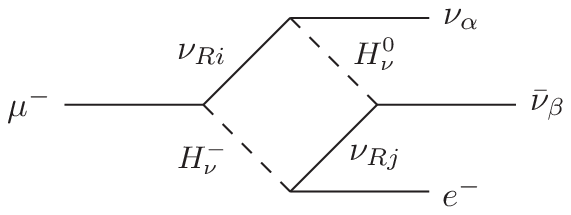}
      &  \includegraphics[width=0.4\textwidth]{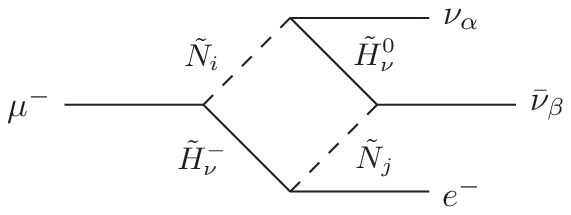}  
  \end{tabular}
  \end{center}
  \caption{The Feynman diagrams of muon decay to electron and left-handed neutrinos via one loop. 
 }
  \label{fig:Muondecay_box}
\end{figure}
\begin{figure}[!t]
  \begin{center}
  \begin{tabular}{c}    
  \includegraphics[width=0.35\textwidth]{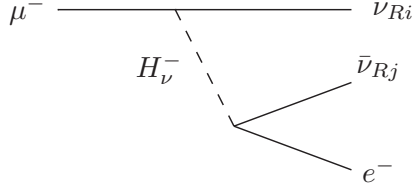}  \end{tabular}
  \end{center}
  \caption{The Feynman diagrams of muon decay to electron and right-handed neutrinos.   
 }
  \label{fig:Muondecay_nuR}
\end{figure}

\subsubsection{loop enhancement of $\mu \rightarrow e \, \bar{\nu}_e \,\nu_{\mu} $}

First, let us estimate the decay width of the main decay mode.
In the following, $p_1,p_2,q_1$ and $q_2$ are the momentum of incoming $\mu$, 
 outgoing $e$, $\nu_{\mu}$ and $\bar{\nu}_e$ respectively.  
The amplitude of $W$ boson mediated process is given by
\begin{equation}
i {\cal M}_1 = \bar{u}(q_1)\frac{g_2}{\sqrt{2}} \gamma^{\mu}P_L u(p_1)\frac{i g_{\mu\nu}}{M_W^2}\bar{u}(p_2) \frac{g_2}{\sqrt{2}}\gamma^{\nu}P_L v(q_2) ,
\end{equation}
 where $g_2$ is the $SU(2)$ gauge coupling and $M_W$ is the $W$ boson mass.
That of RH neutrino and Higgs bosons loop shown in the left window of Fig.~\ref{fig:Muondecay_box} is given by
\begin{eqnarray}
i {\cal M}_2 
 &\simeq & \sum_{i,j} \bar{u}(q_1) \frac{(y_\nu)_{i\mu}}{\sqrt{2}} \gamma_{\mu} (y_\nu)_{i\mu} P_L u(p_1) 
   \bar{u}(p_2) (y_\nu)_{je} \gamma_{\nu}\frac{(y_\nu)_{je}}{\sqrt{2}} P_L v(q_2) \nonumber\\
 && \times \frac{-i}{4(4 \pi )^2}
 ( F_2(M_{H_{\nu}^-},M_{H_{\nu}^0}) + F_2(M_{H_{\nu}^-},M_{A_{\nu}^0}))g_{\mu\nu} ,
\end{eqnarray}
 with the auxiliary function $F_2(x)$, which is defined in the Appendix.
In this estimation of $F_2(x)$,
 we neglect ${\cal O}(M^{-4})$ terms with $M$ being the mass scale of new particles,
 those are much smaller than the leading corrections of ${\cal O}(M^{-2})$.
That of RH sneutrino and Higgsino loop shown in the right window of Fig.~\ref{fig:Muondecay_box} is given by
\begin{eqnarray}
 i {\cal M}_3
 &\simeq&
 \bar{u}(p_2) \gamma_{\mu} P_L u(p_1)\bar{u}(q_1) \gamma_{\nu} P_L v(q_2)g_{\mu\nu}  \nonumber \\
  && \times   \frac{-i \sum_{i, j} (S^Ty_\nu)_{i\mu}^2 (S^T y_\nu)_{je}^2 F_3(M_{\tilde{H}_{\nu}^-},M_{\tilde{H}_{\nu}^0},M_{\tilde{N}_i},M_{\tilde{N}_j})  }{4(4 \pi )^2}. 
\end{eqnarray}
We obtain 
\begin{eqnarray}
 \Gamma(\mu \rightarrow \nu_{\mu} \bar{\nu}_e e)
 &\simeq& \Gamma_{\mu}^{\rm{ (SM) }} \left[
    1 - 2\frac{2M_W^2}{ g_2^2}\frac{\sum_{i, j} (y_{\nu})_{i\mu }^2 (y_{\nu})_{je  }^2 }{8(4 \pi )^2}
 ( F_2(M_{H_{\nu}^-},M_{H_{\nu}^0}) + F_2(M_{H_{\nu}^-},M_{A_{\nu}^0}))
 \right. \nonumber\\ && \left.
   +\frac{1}{2}\frac{2M_W^2}{ g_2^2}\frac{\sum_{i, j} (S^Ty_\nu)_{i\mu}^2 (S^T y_\nu)_{je}^2  F_3(M_{\tilde{H}_{\nu}^-},M_{\tilde{H}_{\nu}^0},M_{\tilde{N}_i},M_{\tilde{N}_j}) }{4(4 \pi )^2}
\right], \label{muondecay:xtra} \\
 \Gamma_{\mu}^{\rm{ (SM) }} &=& \frac{m_\mu^5}{192\pi^3}\frac{1}{2 v^4} , 
\label{muondecay:SM}
\end{eqnarray}
 with the auxiliary function $F_3(x,y,z,w)$, which is given in the Appendix.
Here $m_{\mu}$ is the muon mass and $v\simeq 246\gev$ is the VEV of the SM Higgs field and we keep
 only the leading order loop corrections, namely the interference between tree and one loop.

\subsubsection{$ \mu \rightarrow \nu_{\alpha} \bar{\nu}_{\beta} e $ with $(\alpha, \beta)\neq (\mu, e)$}

Due to the lepton flavor violating neutrino Yukawa coupling,
 the flavor of the final state neutrino can be different from muon-type and anti-electron-type.
However, this decay mode has only loop induced new contribution and is suppressed compared
 to the other contribution to the decay which has the interference term between tree-level and loop induced term.
Thus, this mode is negligible.

\subsubsection{$ \mu \rightarrow \nu_R \bar{\nu}_R e $}

This decay mode with RH neutrinos in the final state is induced by the tree level process
 mediated by the neutrinophilic charged Higgs boson~\cite{0809.5221}.
A worth noting feature is that this is not a $V-A$ interaction but a scalar interaction. 
The amplitude of the process $ \mu \rightarrow \nu_{Ri} \bar{\nu}_{Rj} e $  mediated by $H_{\nu}^-$ shown in the Fig.~\ref{fig:Muondecay_nuR}  is given by
\begin{eqnarray}
i {\cal M}_{H_\nu} &=& \bar{u}(q_1)(y_{\nu})_{\mu i }P_L u(p_1)\frac{i}{M_{H_{\nu}^+}^2}\bar{u}(p_2)(y_{\nu})_{e j} P_R v(q_2) \\
 &\equiv & \frac{4 G_F}{\sqrt{2}}g_{LL}^S \bar{u}(q_1)P_L u(p_1)\bar{u}(p_2) P_R v(q_2).
\end{eqnarray}
Here, we normalise the effective coupling with $G_F$, the Fermi constant measured in the experiment,
 and we introduce a new parameter $g^S_{LL}$ defined by~\cite{Fet} 
\begin{equation}
 g_{LL}^S \equiv \frac{(y_{\nu})_{i\mu  }(y_{\nu})_{je} }{M_{H_{\nu}^+}^2}\frac{\sqrt{2}}{4 G_F}.
\label{def:gLLS}
\end{equation}
The partial width is estimated as
\begin{equation}
 \Gamma(\mu \rightarrow \nu_{Ri} \bar{\nu}_{Rj} e)
 = \frac{\sum_{i,j}|(y_{\nu})_{\mu i }(y_{\nu})_{e j}|^2}{64 M_{H_{\nu}^+}^4} 
   \frac{m_{\mu}^5}{192\pi^2} . \label{muondecay3}
\end{equation}
As we will see later, it turns out
 that this decay mode has to be highly suppressed due to
 the well consistency with the SM. 
Thus, in fact, this would not be significant for muon decay contribution.

\subsubsection{Total}

From Eqs.~(\ref{muondecay:SM}) and~(\ref{muondecay3}),
the final total decay width of muon is given by
\begin{eqnarray}
\frac{ \Gamma_{\mu}}{ \Gamma_{\mu}^{\rm{ (SM) }} }
 &\simeq & \left[
    1 - \frac{\sum_{i, j} (y_{\nu})_{\mu i}^2 (y_{\nu})_{e j }^2 }{8(4 \pi )^2} v^2
 ( F_2(M_{H_{\nu}^-},M_{H_{\nu}^0}) + F_2(M_{H_{\nu}^-},M_{A_{\nu}^0}))
 \right. \nonumber\\ && \left.
   +\frac{v^2 \sum_{i, j} (S^T y_\nu)_{i\mu}^2 (S^T y_\nu)_{je}^2 F_3(M_{\tilde{H}_{\nu}^-},M_{\tilde{H}_{\nu}^0},M_{\tilde{N}_i},M_{\tilde{N}_j}) }{16 (4 \pi )^2}  \right. \nonumber\\
 && \left.+  \frac{v^4\sum_{i,j} (y_{\nu})_{\mu i }^2(y_{\nu})_{e j}^2}{32 M_{H_{\nu}^+}^4} \right].
\label{muondecay:total}
\end{eqnarray}

By comparing  Fermi constant $G_F$ measured from muon decay width and other SM quantities,
 the consistency of the SM can be tested~\cite{Herczeg:1997bu,Marciano:1999ih}.
If we express the Fermi coupling constant with a parameter which stands for a correction
 due to new physics by
\begin{equation}
 G_F = \frac{g_2^2}{4 \sqrt{2} M_W^2 (1-\Delta)},
\end{equation}
 the new physics contribution $\Delta$ is constrained to be~\cite{Erler:2004cx}
\begin{equation}
 \Delta = 0 \pm 0.0006 .
\end{equation}

The last term in Eq.~(\ref{muondecay:total}) comes from the non-($V-A$) interaction
 via $H_{\nu}^+$ mediation.
The muon decay experiments can not measure helicity of produced neutrinos,
 but the inverse decay of muon, $\nu_{\mu} + e \rightarrow \mu + {\rm missing}$,
 well confirms the $V-A$ form interaction and leave a small room for scalar interaction 
 as~\cite{charm}
\begin{equation}
 |g_{\rm LL}^S|^2 < 0.475  \quad ({ \rm 90\% C.L.}).
\label{constr:gLLS}
\end{equation}
With Eq.~(\ref{def:gLLS}), we find the constraint on the $H_{\nu}^+$ mass and Yukawa couplings.

\begin{figure}[htbp]
  \begin{center}
   \includegraphics[width=0.7\textwidth]{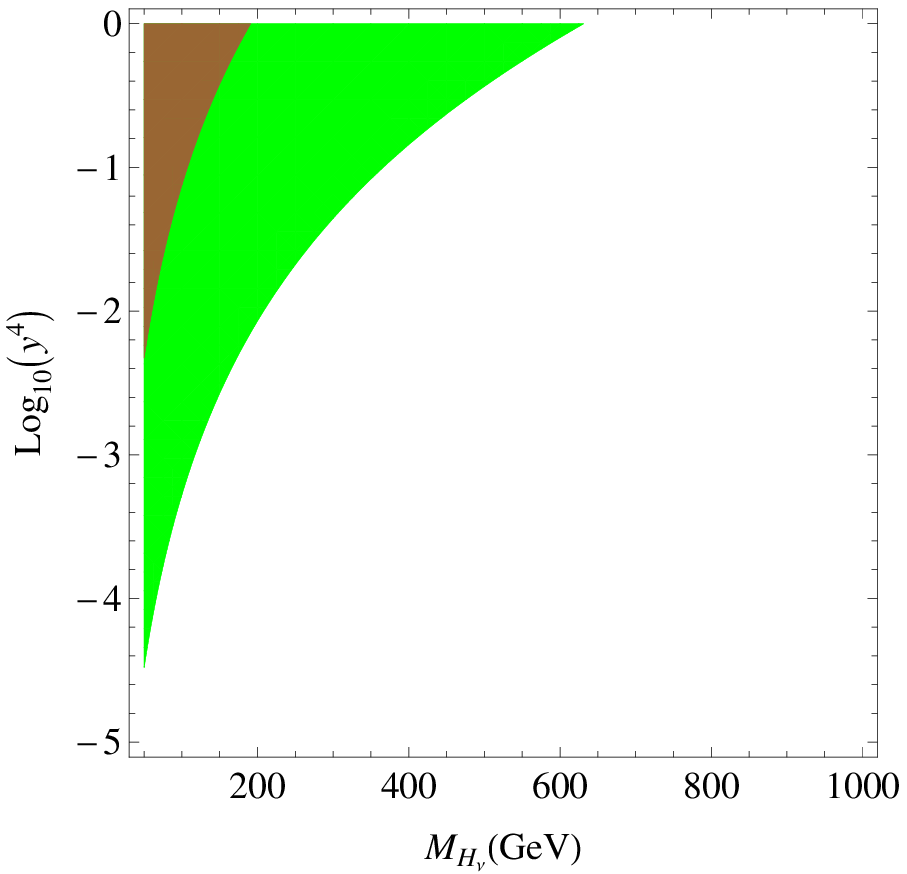}
  \caption{Constraints on charged Higgs boson mass  from the muon decay property. 
  The brown (green) shaded region is excluded by too large $g^S_{LL}$ ($\Delta$). 
} \label{fig:Higgs}
  \end{center}
%
  \begin{center}
  \begin{minipage}{0.48\hsize}
  \begin{center}
   \includegraphics[width=1.0\textwidth]{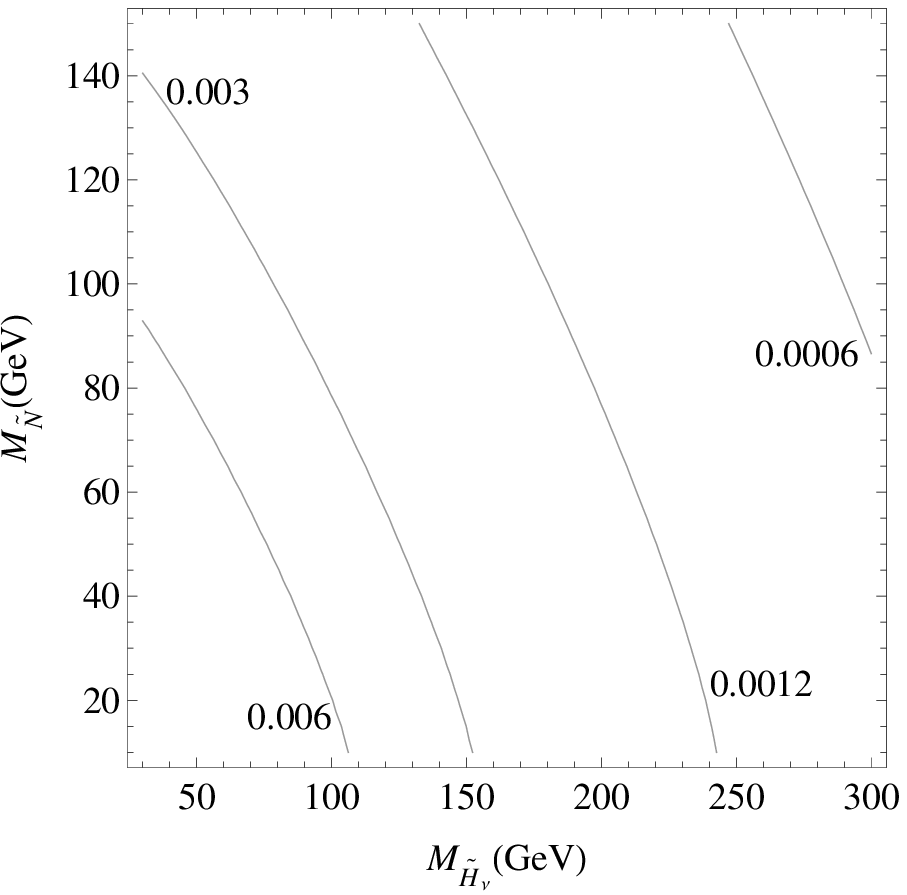}
  \end{center}
  \end{minipage}
 \hfill
  \begin{minipage}{0.48\hsize}
  \begin{center}
   \includegraphics[width=1.0\textwidth]{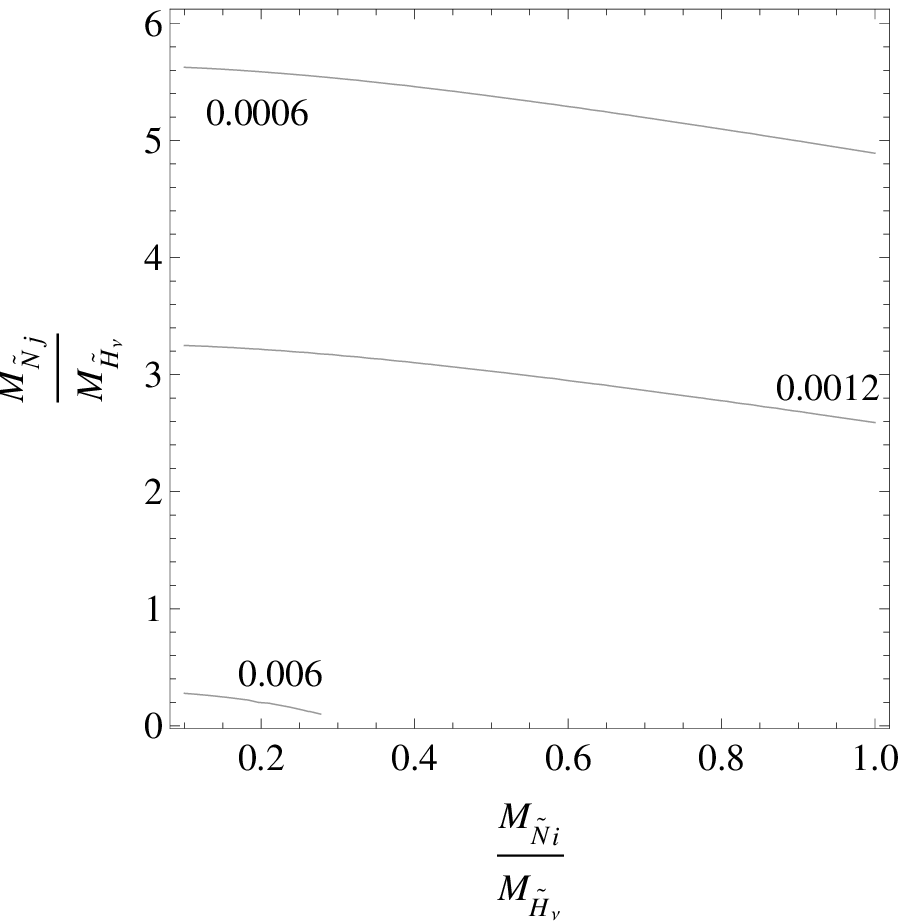}
  \end{center}
  \end{minipage}
  \caption{ Contours of $2\Delta$ of \eq{2Delta} in the plane of
  $(M_{ \tilde{H}_{\nu}^-}, \, M_{ \tilde{N}_i }) $  with $M_{ \tilde{\nu}_i}=M_{ \tilde{N}_j}$ in the left window,
  and  in the plane of $(M_{ \tilde{N}_j }/M_{ \tilde{H}_{\nu}^-},\, M_{ \tilde{N}_j }/M_{ \tilde{H}_{\nu}^-} )$  with $M_{ \tilde{H}_{\nu}^-} = 100$ GeV in the right window. The region $2\Delta \lesssim 0.0012$ is preferred.}
   \label{fig:Snu}
  \end{center}
\end{figure}

First, let us consider the constraints on charged Higgs boson from muon decay
 in the decoupling limit of Higgsino of the second term. 
Then the first term in \eq{muondecay:total} gives dominant contribution to $\Delta$ and the third term to $g^S_{LL}$. 
The constraints on charged Higgs mass and Yukawa coupling
 is shown in Fig.~\ref{fig:Higgs}, where for simplicity we take $M_{ \tilde{H}_{\nu}^-}=M_{ \tilde{H}_{\nu}^0}$.
For Yukawa couplings of the order of unity,
 the mass of charged Higgs must be heavier than around $600\gev$~\footnote{Neutrinophilic Higgs
 bosons with mass {\cal O}(100) GeV is possible
 for $y_{\nu} < {\cal O}(0.01)$~\cite{Davidson:2009ha,Haba:2011nb}. }. 
The cosmological consideration of dark radiation imposes the further stringent
 lower bound on the masses of those extra Higgs bosons, as we will see later in Section~\ref{Constraints:Cosmo}.

Next,
 we need to consider the decoupling limit of very heavy Higgs boson as found just above,
 the the dominant correction comes from the second term in \eq{muondecay:total} 
 from the sneutrino and chargino loop, namely
\begin{equation}
2\Delta\simeq \frac{v^2 F_3(M_{\tilde{H}_{\nu}^-},M_{\tilde{H}_{\nu}^0},M_{\tilde{N}_i},M_{\tilde{N}_j}) }{16 (4 \pi )^2}.
\label{2Delta}
\end{equation}
In Fig.~\ref{fig:Snu},
we show the contour plot of 
$2\Delta$ for ${\cal O}(1)$ Yukawa couplings
 in the plane of $(M_{ \tilde{H}_{\nu}^-}, \,M_{ \tilde{N}_i })$
 with $M_{ \tilde{N}_i}=M_{ \tilde{N}_j}$ in the left window ,
 and that in the plane of $(M_{ \tilde{N}_i }/M_{ \tilde{H}_{\nu}^-},\,  M_{ \tilde{N}_j }/M_{ \tilde{H}_{\nu}^-} )$ 
 with $M_{ \tilde{H}_{\nu}^-} = 100$ GeV in the right window.
We can see that the charged Higgsino $\tilde{H}_{\nu}^-$ need to be heavier than a few hundreds GeV
 for degenerate sneutrino case, $M_{ \tilde{N}_j } \simeq M_{ \tilde{N}_i }$,
 or
 two of RH sneutrinos need to be several times heavier than chargino with $100$ GeV mass
 and only one RH neutrino can be light.

\subsection{Collider constraints}
In our model,
 the neutrinophilic Higgsinos  $(\tilde H_\nu^0,\, \tilde H_\nu^+$) are $SU(2)$ doublet and
 can be light enough to be produced at the on-shell from $pp$ or $e^+e^-$ collisions, and subsequently decay 
to leptons ($l_i$) and lightest sneutrino ($\tilde N_{DM}$), which similar to the production of Wino/Zino and subsequent decay to the lightest neutralino in the minimal supersymmetric standard model (MSSM).
However, the collider constraints on our model with the RH sneutrino LSP is
 slightly different from the current searches on SUSY based on the neutralino LSP. 
 
The production channels of the neutrinophilic Higgsinos are $s$-channel $Z^0/\gamma$ boson exchange for $e^+e^-$, 
$pp$ collisions, $t$-channel $\tilde N$ exchange for $e^+e^-$ collision, and $s$-channel $W^\pm$ exchange 
for $pp$ collision. They subsequently decay 
to leptons ($l_i$) and the lightest sneutrino ($\tilde N_{DM}$),
\dis{
&\tilde H^+_\nu\to \tilde N_i l^+_i, \quad 
\tilde H^0_\nu\to \tilde N_i\bar\nu_j,\\
&\tilde N_i^* \to l_i^+  \bar{\tilde H}^+_\nu\, (\bar\nu_j \bar{\tilde H}^0_\nu)\ 
{\rm or}\  l_i^+ e^- \tilde N_{DM}^*\, (\bar\nu_j\nu_k \tilde N_{DM}^*)\quad {\rm for}\ \ i\neq {\rm DM},
}
where $\nu_i$ is the mass eigenstate of neutrino, and
 we assume that $\tilde N_{DM}$ is the lightest RH sneutrino as dark matter.
For a case where those particle are too heavy to be produced at the on-shell,
 the production cross section is kinematically very suppressed. 
For cases that RH sneutrinos are light but degenerate,
 only three-body decay is possible via virtual Higgsino ($\tilde{H}_\nu$) and
 the energy of the produced lepton is much suppressed. 

\begin{figure}[!t]
  \begin{center}
  \begin{tabular}{cc}    
  \includegraphics[width=0.4\textwidth]{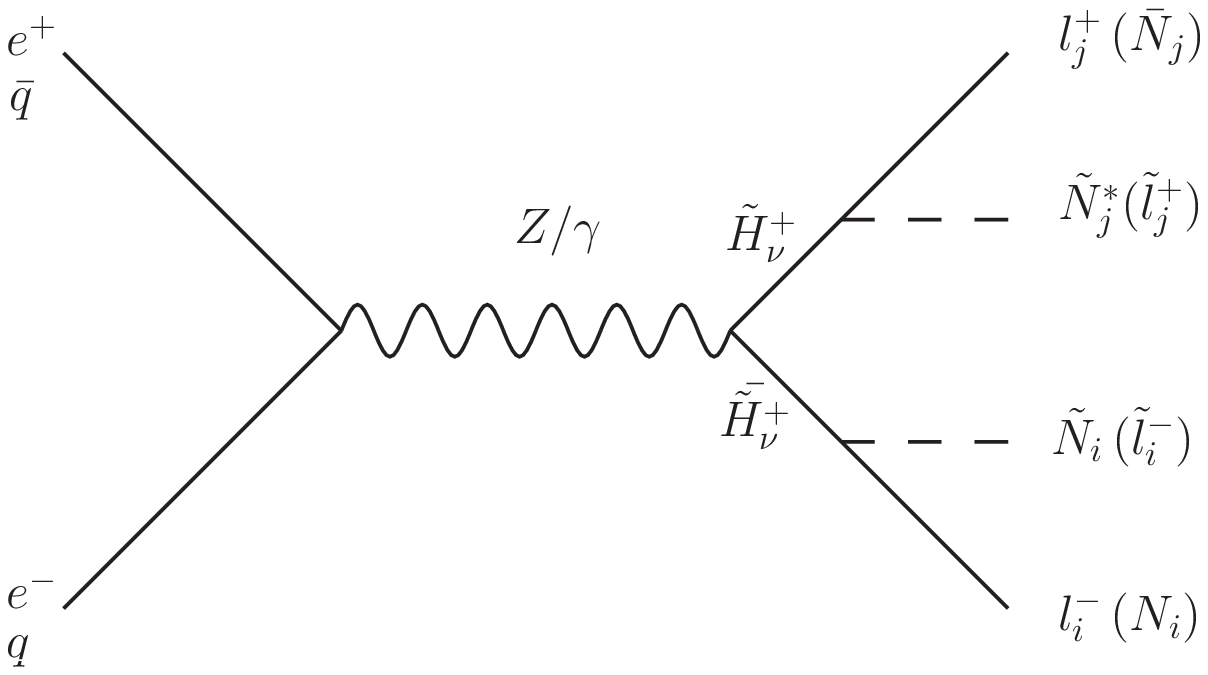}
      &  \includegraphics[width=0.4\textwidth]{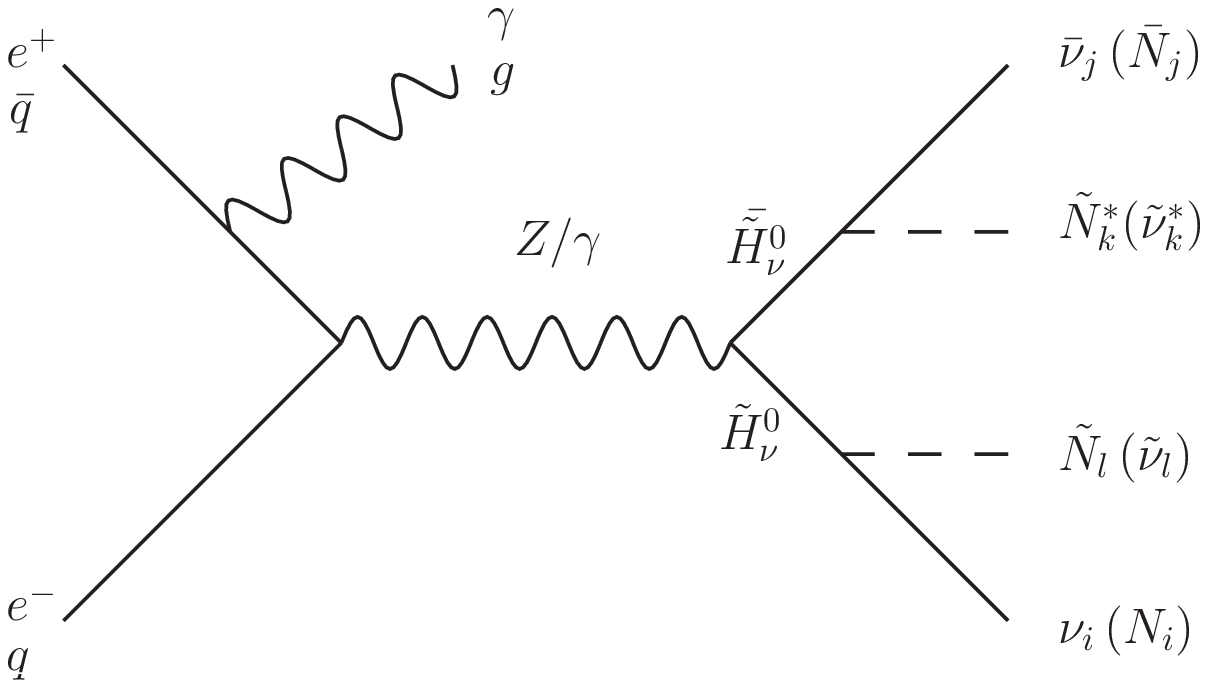}   \\
   \hskip -0.5cm  \includegraphics[width=0.32\textwidth]{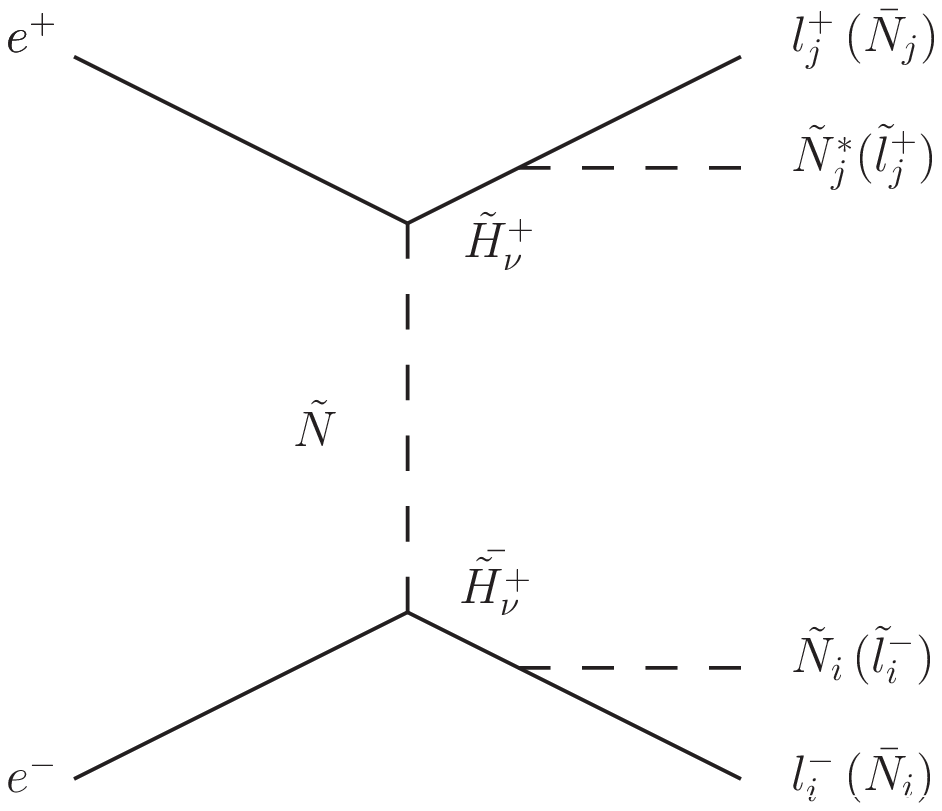}
   &
    \includegraphics[width=0.42\textwidth]{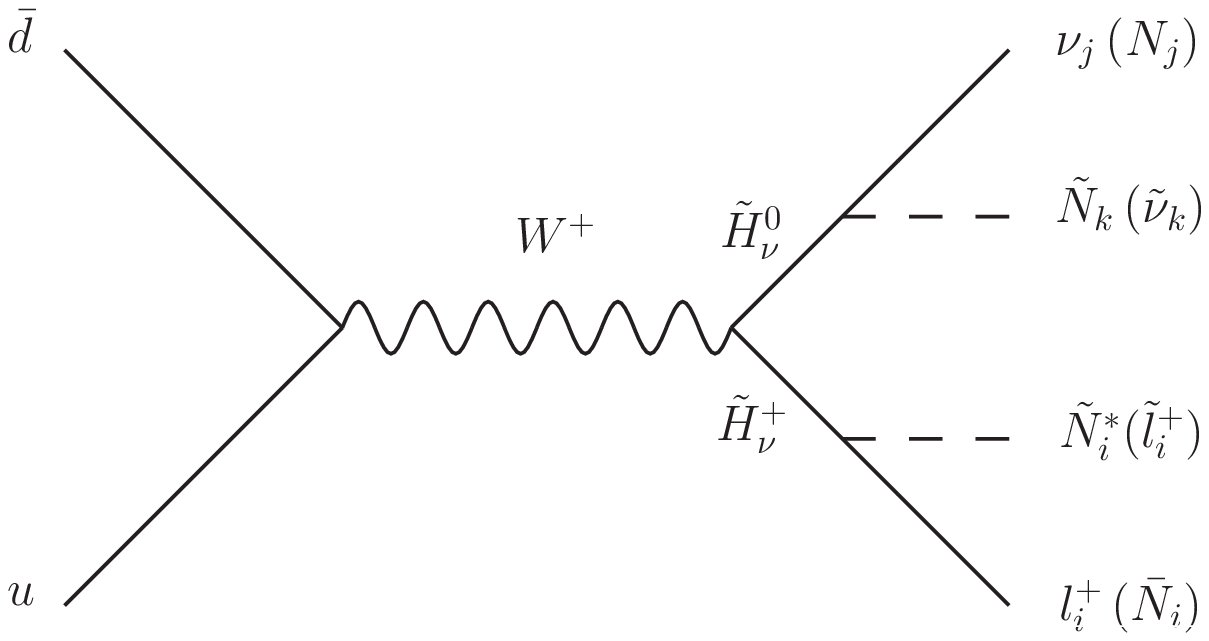}
  \end{tabular}
  \end{center}
  \caption{The neutrinophilic Higgsinos ($\tilde H_\nu$) production and its sequent decay into RH sneutrino. 
  $\tilde H_\nu$s are produced by the $s$-channel of $Z^0/\gamma$, $t$-channel of $\tilde N$, and 
  $s$-channel of $W^\pm$ exchanges. 
   }
  \label{fig:higgsinoproduction}
%
%
  \begin{center}
  \begin{tabular}{c}
   \includegraphics[width=1\textwidth]{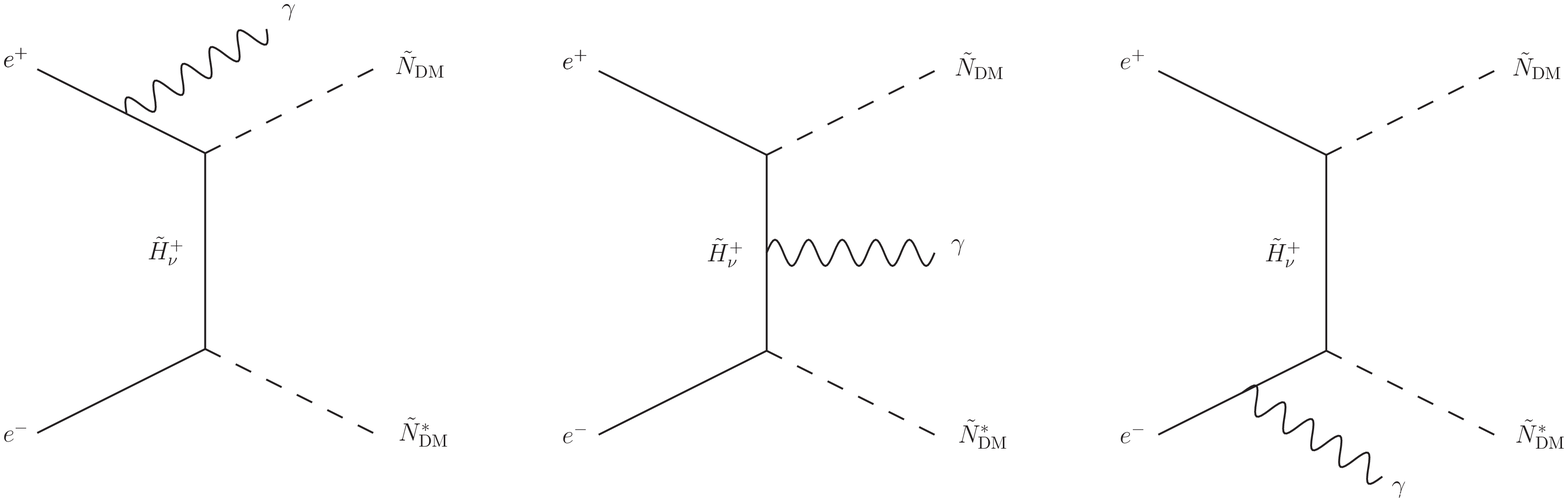}
     \end{tabular}
  \end{center}
  \caption{Feynman  diagrams for $e^+e^-\to\tilde N_{DM}\tilde N_{DM} \gamma$.
  Since the produced RH sneutrinos are the lightest superpartners, $\tilde N_{DM}$, the only observable 
  signal is photon.}
  \label{fig:monophoton_LEP}
\end{figure}

The corresponding diagrams are given in Fig.~\ref{fig:higgsinoproduction}.
The final decay products are multi leptons plus a large missing energy by $\tilde{N}_{DM}$,
\dis{
e^+e^-(q\bar q)&\to  l^+_i l^-_j+{\rm missing\, E}+(n\, l^+l^-),\\
u\bar d &\to l^+_i+{\rm missing\, E}+ (n\,  l^+l^-),
}
 with $n$ being an integer,
or mono photon plus a large missing energy by $\tilde{N}_{DM}$,
\dis{
e^+e^-(q\bar q)&\to  \gamma  +{\rm missing\, E}.
}

\subsubsection{Constraints from LEP}

As far as the LEP bound on chargino is concerned,
 if the mass of the Higgsinos are greater than the threshold energy scale for 
$e^+e^-$ collision ($\sqrt{s}/2\simeq 104\,{\rm GeV}$),
 it is natural to expect that the constraints are relaxed drastically. 
Thus, we take $m_{\tilde{H}_\nu} \gtrsim 103.5\gev$~\cite{LEPSUSY} as the kinematical lower bound
 to avoid the LEP searches for direct production of charginos. 
 
For the direct production of $\tilde N_i$, the $t$-channel $\tilde H_\nu$ exchange for $e^+e^-$ collision 
is dominant as in FIG. \ref{fig:monophoton_LEP}. 
\dis{
e^+e^-\to \gamma+{\rm missing\, E}. \label{lepmonophoton}
}
In this case, the result of monophoton searches for the MSSM neutralino $e^+e^-\to \tilde\chi^0_1\tilde\chi^0_1\gamma$ can be used to constrain the mass and couplings of the RH sneutrino to electrons~\cite{Fox:2011fx}. In this analysis, using the effective operator  $\bar{\chi}e \bar{e}\chi/\Lambda $, the cutoff scale $\Lambda$  should be greater than about $330$\,{\rm GeV} for the fermion dark matter mass $m_{\chi} < 80$ GeV as shown in figure 7.

\begin{figure}[!t]
  \begin{center}
  \begin{tabular}{c}
   \includegraphics[width=0.7\textwidth]{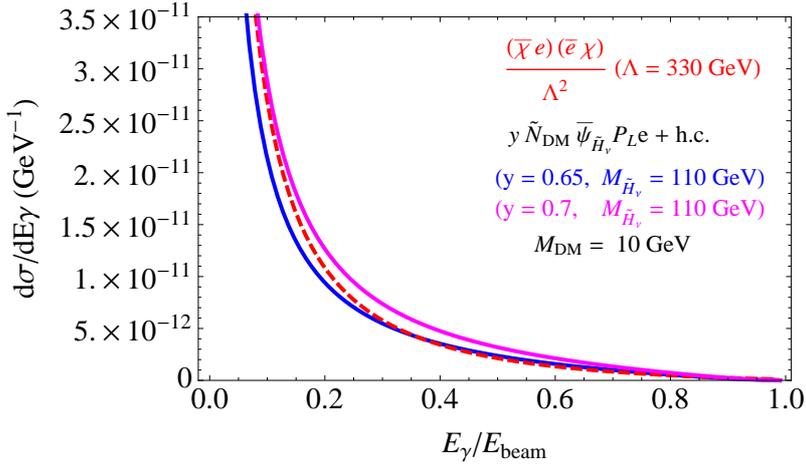}
     \end{tabular}
  \end{center}
  \caption{ Production rate ($d\sigma/dE_\gamma$)  with respect to 
  $E_\gamma/E_{\rm beam}$  for ($e^+e^-\to\gamma$ plus DM pair) in two models.
  Red dashed line corresponds to the model studied in Ref.~\cite{Fox:2011fx}.
  The studied process is ($e^+e^-\to \gamma\chi\bar\chi$) from the interaction term 
  $(\bar\chi e)(\bar e \chi)/\Lambda^2$. 
  Depending on the dark matter mass $M_{\rm DM}$, the 
  the lower limit of $\Lambda$ is different. 
  The DELPHI(LEP) monophoton search for $M_{\rm DM}=10$ GeV
  gives the lower bound on $\Lambda \geq 330$ GeV.
  The second model is ours.  
  The event is ($e^+e^-\to \gamma \tilde N_e\tilde N_e^*$) from the interaction term
  $y \tilde N_{DM}\bar \psi_{\tilde H_\nu} P_L e + h.c.$. 
    We take $m_{\tilde H_\nu}=110\, {\rm GeV}$, and 
  the cases with two Yukawa couplings ($y=0.65,\, 0.7$)
  are presented (Blue and Magenta).
  $E_{\rm beam}$ is taken as $100$ GeV, which is the average value for the LEP search. 
  We find that our model parameters ($y$ and $m_{\tilde H_\nu}$) are 
  constrained by the LEP search as did in Ref.~\cite{Fox:2011fx}, i.e. 
  the region above the red dashed line is ruled out. 
  }
  \label{fig:e+e-togamma}	
\end{figure}

\subsubsection{LHC bound}

The current 8 TeV LHC gives lower bound on the charged Higgsino mass of $550\gev$,
 leaving the degenerate region with $M_{\tilde H}-M_{\tilde N}<50 \gev$ unconstrained
 from the chargino decay into light leptons, namely $e$ or $\mu$.
For the decay into $\tau$ lepton,
 the bound is relaxed and requires $M_{\tilde H} \gtrsim 350\gev$ and $M_{\tilde N} \gtrsim 150\gev$~\cite{Guo:2013asa}.

\subsection{lepton flavor violation ($\mu \rightarrow e \gamma $)}

\begin{figure}[!t]
  \begin{center}
  \begin{tabular}{ccc}    
  \includegraphics[width=0.28\textwidth]{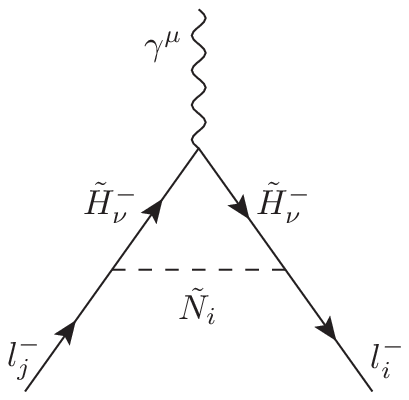}
  &
      &  \includegraphics[width=0.28\textwidth]{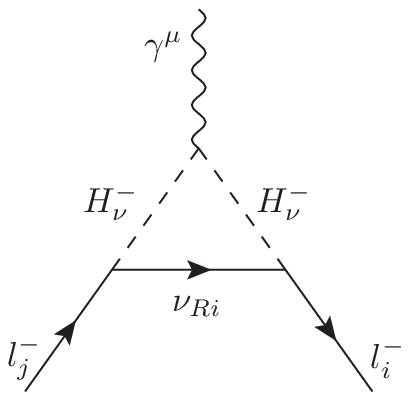}  
  \end{tabular}
  \end{center}
  \caption{The Feynman diagrams for the lepton flavor violation corresponding to \eq{LFVamp}.
 }
  \label{fig:FCNC}
\end{figure}
The off-diagonal components of Yukawa couplings $(y_{\nu})_{i \alpha}$ in \eq{Yukawa} induce
 lepton flavor violating (LFV) decay of leptons.
The general effective operator can be written as
\begin{equation}
{\cal M}=e {\epsilon^{\mu}}^* \overline{u_i}(p-q)\Big[im_j\sigma_{\mu\nu}q^\nu (A_2)_{ij} + i m_j \sigma_{\mu\nu} q^\nu \gamma_5 (A_3)_{ij}  \Big]u_j(p),
\label{LFVamp}
\end{equation}
with $\sigma_{\mu\nu} = \frac{i}{2}[\gamma_\mu,\gamma_\nu]$.
The decay rate of $l_j \rightarrow l_i \gamma$ is given by
\begin{equation}
\Gamma(l_j\to l_i\gamma)=\frac{e^2}{8\pi}m_{l_j}^5 (|(A_2)_{ij}|^2+|(A_3)_{ij}|^2).
\label{LFVwidth}
\end{equation}
For the muon decay, 
$\mu \to e\gamma$, the branching ratio is given by
\begin{equation}
{\rm Br}(\mu \to e\gamma)=\frac{96\pi^3 \alpha}{G_F^2}(|(A_2)_{12}|^2+|(A_3)_{12}|^2).
\end{equation}
The present bounds on the branching ratios of the LFV decays are~\cite{Adam:2013mnn,Aubert:2009ag}
\dis{
&{\rm Br}(\mu\rightarrow e \gamma ) <  5.7\times 10^{-13 }\quad ({ \rm 90\% C.L.}),\\
&{\rm Br}(\tau\rightarrow e \gamma ) <  3.3\times 10^{-8 } \quad ({\rm 90\% C.L.}),\\
&{\rm Br}(\tau\rightarrow \mu \gamma ) <  4.4\times 10^{-8 }\quad ({\rm 90\% C.L.}).
\label{LFVconstraints}
}

For our model with Yukawa interactions in \eq{Yukawa}, from the diagrams given in figure 8, we obtain
\dis{
(A_2)_{\alpha \beta} & \simeq
 \frac{1}{32\pi^2}  \left(  \sum_{l}
 \frac{ ( S^T y_\nu)_{l\alpha} (S^T y_\nu)_{l\beta}}{M_{\tilde{H}_\nu^-}^2}
 F\left(\frac{M_{\tilde{N}_{l}}^2}{M_{\tilde{H}_\nu^-}^2}\right)
 - \sum_k \frac{ (y_{\nu})_{k \alpha } (y_{\nu})_{k\beta} }{48 M_{H_\nu^-}^2}
  \right) ,  \quad \\
(A_3)_{\alpha \beta} &= 0,
}
 with $F(x)$ being an auxiliary function.
The experimental limits~(\ref{LFVconstraints}) give strong bounds on Yukawa couplings as well as masses of 
mediated particle, $M_{\tilde{N}_{k}}, M_{\tilde{H}_\nu^-}$ and $M_{H_\nu^\pm}$, as we will show.
In fact, for ${\cal O}(100)$ GeV masses of sneutrinos and the neutrinophilic chargino,
 we find the LFV decaying branching ratios are of ${\cal O}(10^{-6})$, those are very large compared
 with current bounds~(\ref{LFVconstraints}).
Since Yukawa couplings are fixed from the neutrino mass and are the order of unity,
 the charged Higgs must be heavier than around $10$ TeV in the second term. 
On the other hand, for the first term, we have a possibility that
 one of sneutrinos and charged Higgsinos are relatively light
 in the case that $\tilde y = S^T y_\nu$ are
 suitably aligned by appropriate sneutrino mixings $\sigma_{ij}$
 in such a way that the flavor-violating processes are suppressed enough.
Such mixing angles can be found by requiring some off-diagonal components of
 $\tilde{y} = S^T y_\nu$ to be almost vanishing.

\subsection{Anomalous magnetic dipole moment of muon}

The muon anomalous magnetic dipole moment has $3.3\,\sigma$ discrepancy
 between experimental data and the SM prediction as~\cite{ Bennett:2006fi,Hagiwara:2011af}
\begin{equation}
a_\mu({\rm EXP}) - a_\mu ({\rm SM}) = (26.1\pm8.0)\times 10^{-10}. \label{deltaamu}
\end{equation}
Thus, this discrepancy has been regarded as a hint and provided a motivation
 to investigate new physics beyond the standard model of particle physics.

$A_2$ and $A_3$ in Eq.~(\ref{LFVamp}) contribute to the magnetic and electric dipole moment, respectively.
The resultant electric and magnetic dipole moment of $l_j$ lepton are given as
\begin{eqnarray}
d_{l_j} &=& m_{l_j}(A_3)_{jj},\\
 a_{l_j} &=& \frac{(g-2)_{l_j}}{2} =2m_{l_j}^2(A_2)_{jj}.
\end{eqnarray}
The additional contribution to the induced magnetic moment of muon
 in the supersymmetric neutrinophilic Higggs model with large Yukawa couplings is given by 
\begin{equation}
\delta a_{\mu} = 2 m_{\mu}^2\frac{1}{32\pi^2} \sum_l 
   \frac{(S^Ty_\nu)_{l\mu} (S^Ty_\nu)_{l\mu}  }{M_{\tilde{H}^-}^2}
   F\left(\frac{M_{\tilde{\nu}_l}^2}{M_{\tilde{H}^-}^2}
  \right) ,
\label{amu}
\end{equation}
 where we assumed Yukawa couplings are real and the negligible charged Higgs boson contribution is omitted.
We might expect a large $g-2$ of the muon for light sneutrinos
 and the light $\tilde{H}_{\nu}$-like chargino
 because Yukawa coupling constants are ${\cal O}(1)$.
However, as mentioned above,
 the experimental limits on the lepton flavor violation in~\eq{LFVconstraints}
 are very stringent and we need a special mixing of sneutrinos.

\subsection{Compatibility in benchmarks}

As mentioned previous subsections, LFV constraints are very stringent.
Indeed, for cases of vanishing lightest neutrino mass $m_1$
 as in benchmark point 1 and 3 in Table~\ref{tab:benchmarks},
 we could not find viable parameter sets.
Thus, here we mention viable parameter sets based on
 the benchmark point 2 and 3 in the Table~\ref{tab:benchmarks}. 

At first, for the benchmark 2,
we find that LFV constraints are avoided with the sneutrino mixing angles 
 $(\sigma_{12}, \sigma_{23},\sigma_{13}) \simeq (0.75, 0.68, 0.20)$,
 and the resultant $\tilde{y}$ is given by 
\begin{eqnarray}
 \tilde{y} = S^T y_\nu \simeq 
\left( 
\begin{array}{ccc}
1.18 & 0.06 & 0 \\
0 & 1.28 & 0 \\
0.05 & 0.24 & 1.31 
\end{array} 
\right) .
\end{eqnarray}
Then, the mass of the lightest RH sneutrino $\tilde{N}_2$ can be $10-100$ GeV
 without inducing a large LFV,
 while the mass of the other two $\tilde{N}_1$ and $\tilde{N}_3$ need to be ${\cal O}(1-10)$ TeV. 
Notice that with the definition of $S$ by Eq.~(\ref{def:sneu:masseigenstate}), 
 we take the mass ordering of RH neutrinos
 as $M_{\tilde{N}_2} < M_{\tilde{N}_1} < M_{\tilde{N}_3}$.
With this choice of sneutrino mixing angles, the muon decay width constraint
 discussed Sec.~\ref{muondecay} is avoided. 
Here, we can see that the coupling between the lightest RH sneutrino ($\tilde{N}_2$) and
 the electron is negligibly small. 
This small coupling automatically suppresses the LEP mono-photon constraint from \eq{lepmonophoton}.
 \begin{figure}[!t]
  \begin{center}
  \begin{tabular}{cc}    
  \includegraphics[width=0.45\textwidth]{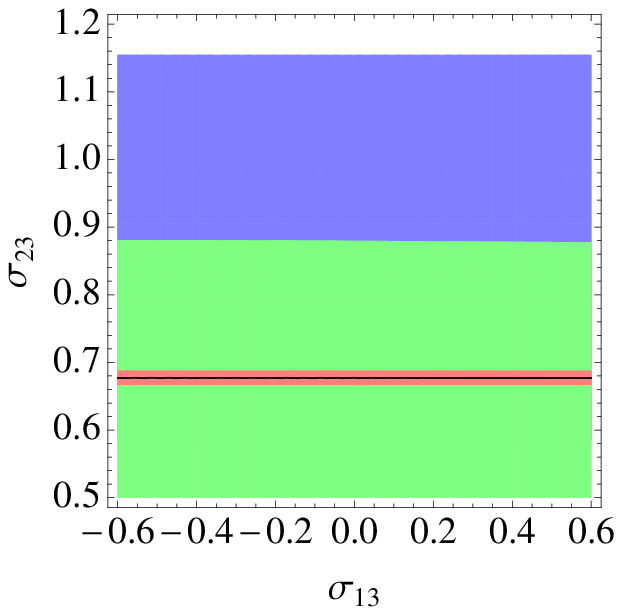}
      &  
      \includegraphics[width=0.45\textwidth]{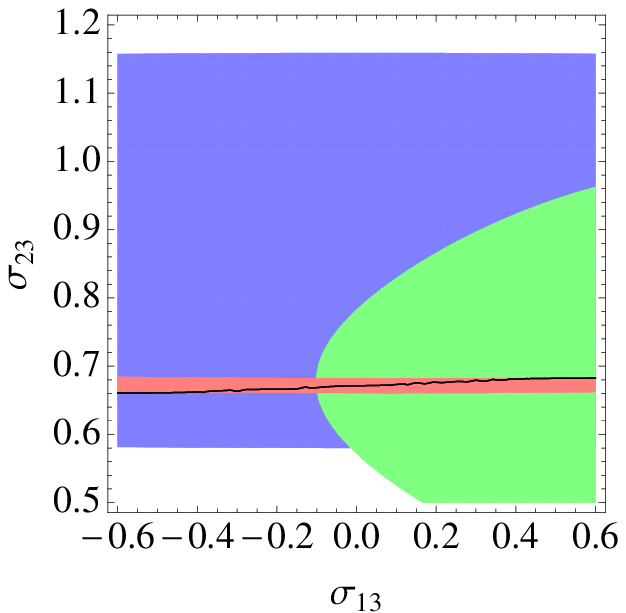}   \\
   \hskip -0.1cm  
   \includegraphics[width=0.45\textwidth]{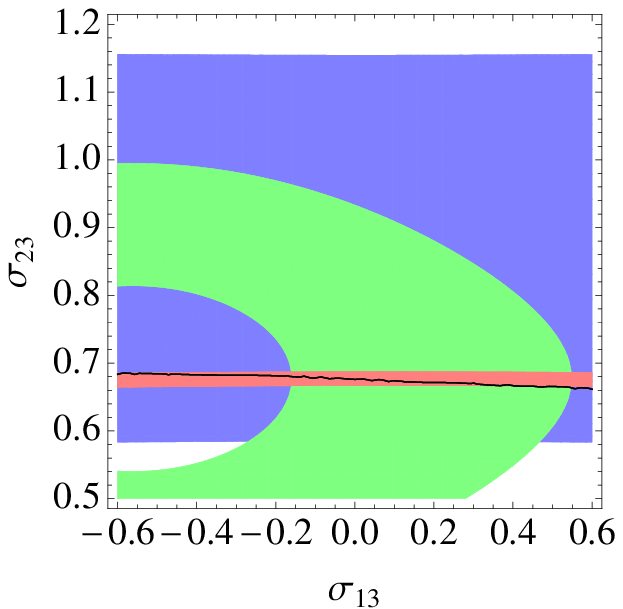}
   &
    \includegraphics[width=0.45\textwidth]{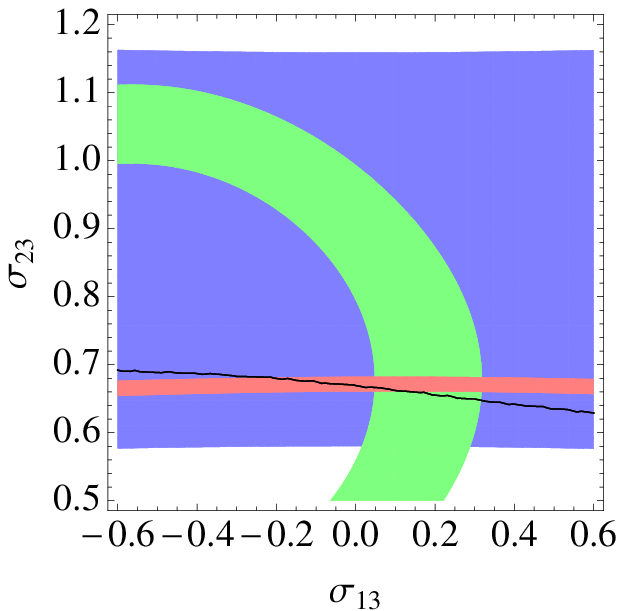}
  \end{tabular}
  \end{center}
  \caption{Allowed parameter space in the plane of the sneutrino 
  mixing angles ($\sigma_{13}$, $\sigma_{23}$) from the constraints on LFV 
  for the benchmark 2. For all plots, we fixed $M_{\tilde H_\nu}=110$ GeV, 
  $M_{\tilde N_2}=60$ GeV and $\sigma_{12}=0.75$. 
  Top left corresponds to $(M_{\tilde N_1}, M_{\tilde N_3})=(7$ TeV$, 10$ TeV).
  Top right, ($7$ TeV, $1$ TeV). 
  Bottom left, ($1$ TeV, $7$ TeV). 
  Bottom right, ($0.5$ TeV, $1$ TeV).
  The black region is ${\rm Br}(\mu\to e\gamma) < 5.7 \times 10^{-13}$,
  the red region is ${\rm Br}(\tau\to\mu \gamma) < 4.4 \times 10^{-8}$, and 
  the green region is ${\rm Br}(\tau\to e\gamma) < 3.3\times 10^{-8}$.
  The blue region denotes $\delta a_\mu > 10^{-9}$.
 }
  \label{fig:amuLFVben2}
 \end{figure}
 
In Fig.~\ref{fig:amuLFVben2}, we show the viable sneutrino mixing angles from the 
 constrains on the LFV processes for the benchmark 2. 
In those plots, we fixed one mixing angle $\sigma_{12}=0.75$, 
 the neutrinophilic Higgsino mass $M_{\tilde H_\nu}=110$ GeV, and 
 the lightest sneutrino mass $M_{\tilde N_2}=60$ GeV. 
Heavy sneutrino masses, 
 $(M_{\tilde N_1}, N_{\tilde N_3})$ are taken from around TeV to $10$ TeV 
As the figures show, the constraint from $\mu\to e \gamma$ is most serious, and 
 very small regions are allowed. 
Including the constrains from $\tau\to \mu\gamma$, we find that the value of $\sigma_{23}$ should be around $0.68$, which corresponds to $(S^T y_\nu)_{2e}\approx 0$. 
The constraints do not restrict the value of $\sigma_{13}$ much for the case with two heavy sneutrino masses. 
In the allowed parameter space, the sizable $\delta a_\mu$ can be obtained as denoted by the blue colored region.

\begin{figure}[!t]
  \begin{center}
  \begin{tabular}{c}
   \includegraphics[width=0.8\textwidth]{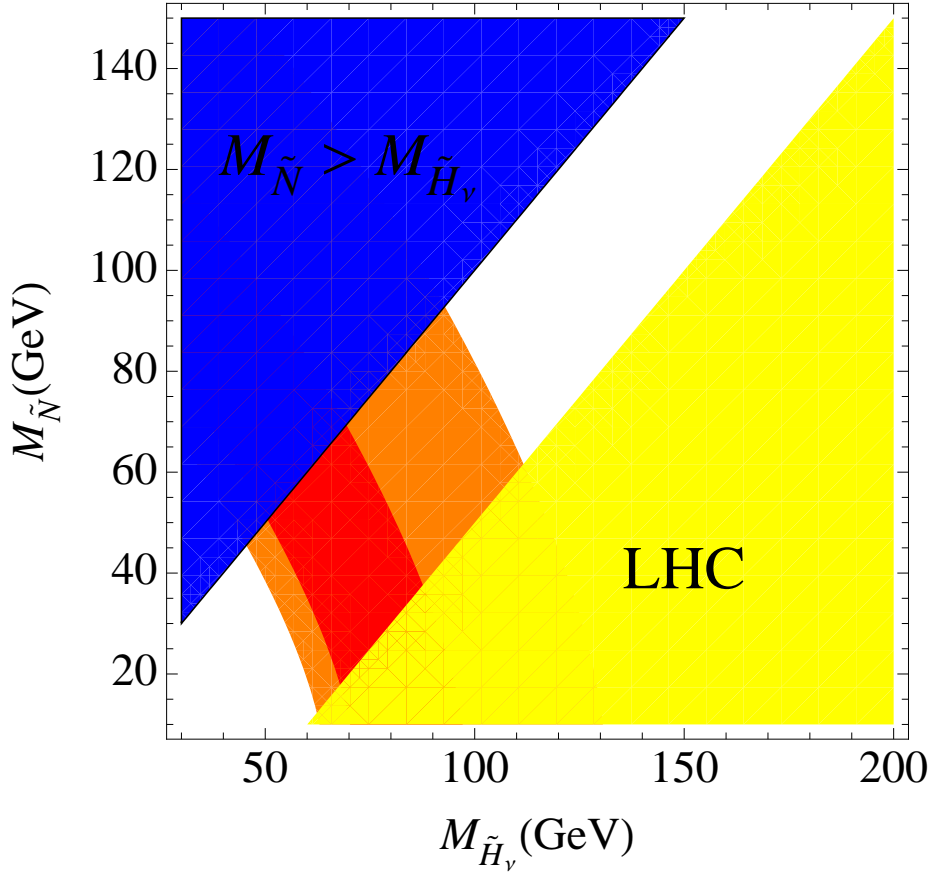}
  \end{tabular}
  \end{center}
  \caption{The viable parameter space for the benchmark 2
 with tuned sneutrino mixing angles so that LFV processes are sufficiently suppressed.
Contours show the contribution to the muon anomalous magnetic moment from \eq{amu} in the plane of the mass of neutrinophilic Higgsino and the RH sneutrino mass $M_{\tilde{N}}$ . 
 $1\sigma$ and $2\sigma$ region of \eq{deltaamu} is shown with Red and Orange color respectively. 
In the blue region, $M_{{\tilde N}} > M_{\tilde H_{\nu}}$ is realized. 
The yellow region, where the mass splitting between sneutrino and chargino is too large, is constrained by the LHC results.}
\label{fig:amu2}
\end{figure}

In Fig.~\ref{fig:amu2}, 
 we show the viable parameter space for the benchmark 2 with contours of
 the contribution to the muon anomalous magnetic moment from \eq{amu}
 in the plane of the mass of neutrinophilic Higgsino and the RH sneutrino mass $M_{\tilde{N}_2}$,
 which is the lightest supersymmetric particle.
Here, we use $M_{\tilde{N}_1} = 7$ TeV and $M_{\tilde{N}_3} = 10$ TeV for reference.
The region of Red and Orange color respectively show $1\sigma$ and $2\sigma$ range of \eq{deltaamu}. 
The blue region corresponds $m_{{\tilde N}_2} > m_{\tilde H_{\nu}}$. 
The yellow region, where the mass splitting between sneutrino and chargino is too large, is constrained by the LHC results.

\begin{figure}[!t]
  \begin{center}
  \begin{tabular}{c}
   \includegraphics[width=0.8\textwidth]{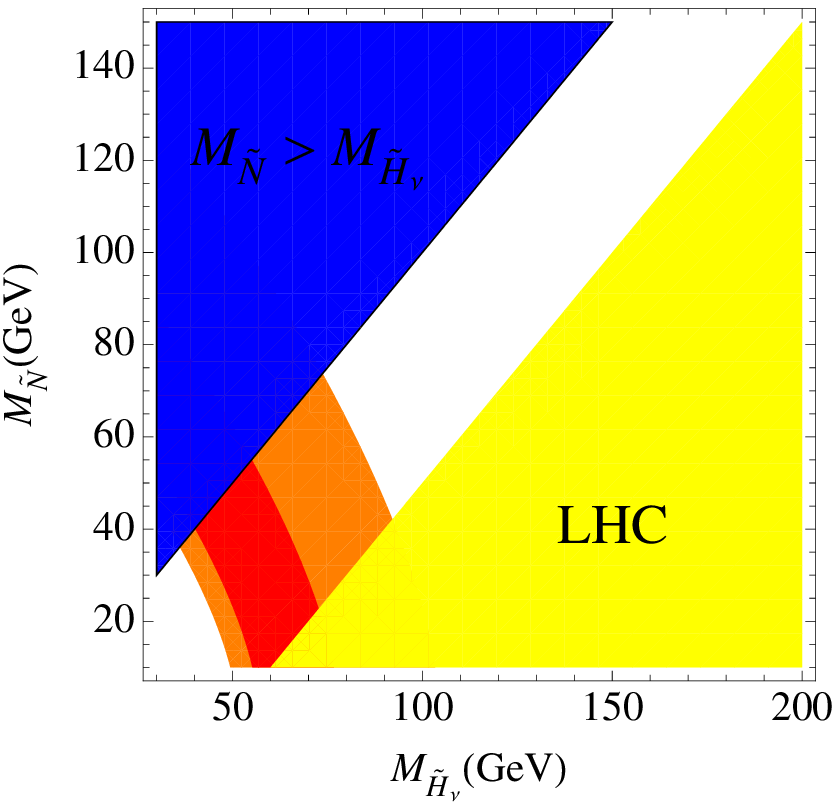}
  \end{tabular}
  \end{center}
  \caption{ Same as Fig.~\ref{fig:amu2} for the benchmark 3.
 }
  \label{fig:amu3}
\end{figure}

For the benchmark 3,
 if the sneutrino mixing angles are
 $(\sigma_{12}, \sigma_{23},\sigma_{13}) \simeq (0.75, 0.93, 0.02)$,
 the resultant $\tilde{y}$ is given by 
\begin{eqnarray}
 \tilde{y} = S^T y_\nu \simeq 
\left( 
\begin{array}{ccc}
1.20 & 0 & -0.25 \\
0 & 1.02 & 0 \\
0.21 & -0.33 & 1.05 
\end{array} 
\right) .
\end{eqnarray}
Then, the mass of the lightest RH sneutrino $\tilde{N}_2$ can be $10-100$ GeV
 without inducing a large LFV,
 while the mass of the other two $\tilde{N}_1$ and $\tilde{N}_3$ need to be ${\cal O}(1-10)$ TeV. 
In Fig.~\ref{fig:amu3}, we show the viable region with $M_{\tilde{N}_1} = 7$ TeV and $M_{\tilde{N}_3} = 8 $ TeV.
Again, we take the mass ordering of RH sneutrinos
 as $M_{\tilde{N}_2} < M_{\tilde{N}_1} < M_{\tilde{N}_3}$.


\section{Cosmological constraints}
\label{Constraints:Cosmo}

\subsection{dark radiation}
\label{DR constr.}

RH component of neutrinos could contribute to the additional relativistic degrees of freedom
 in the early Universe, which is constrained by big bang nucleosysthesis and
 cosmic microwave background radiation observation as $\Delta N_{eff} \lesssim 0.57$.
In our model, RH neutrinos could be in thermal equilibrium due to the scatterings with
 charged leptons (left-handed neutrinos) 
 through neutrinophilic charged (neutral) Higgs bosons via the Yukawa interaction in \eq{Yukawa}.

In order to suppress $\Delta N_{eff}$ enough, 
 RH neutrinos should have decoupled from the thermal bath much before the quark-hadron phase transition
 which takes place at the cosmic temperature $T_c\simeq 200$ MeV.

\subsubsection{$l^-l^+ \longleftrightarrow \nu_R \bar{\nu_R} $}

The scatterings between charged lepton and RH neutrinos are mediated
 by the neutrinophilic charged Higgs boson $H_{\nu}^\pm$.
We obtain
\begin{equation}
\int \overline{|{\cal M}|^2} d{\rm LIPS} \simeq  \frac{1}{8\pi}\frac{|y_\nu y_\nu|^2}{M_{H_{\nu}^+}^4} \frac{s^2}{12} ,
\end{equation}
 with $s$ being the energy at the center of mass frame. Taking thermal average, we find
\begin{equation}
\langle \sigma v \rangle \simeq \frac{|y_\nu y_\nu|^2}{32\pi M_{H_{\nu}^+}^4} T^2 .
\end{equation}

\subsubsection{$ \nu_L \bar{\nu_L} \longleftrightarrow \nu_R \bar{\nu_R} $}

Similarly, the thermal scattering cross section between left-handed and right-handed components
 of neutrinos via $H_{\nu}^0$ and $A_{\nu}^0$ is estimated as 
\begin{equation}
\langle \sigma v \rangle \simeq \frac{|y_\nu y_\nu|^2}{32\pi}\frac{1}{4}
   \left(\frac{1}{M_{H_{\nu}^0}^2} + \frac{1}{M^2_{A_{\nu}^0}} \right)^2 T^2 .
\end{equation}

\subsubsection{Decoupling condition}

The decoupling condition of RH neutrino at the quark-hadron transition epoch is expressed as
\begin{eqnarray}
 \langle \sigma v \rangle n |_{T_{QH}} < H|_{T_{QH}},
\end{eqnarray}
 which is rewritten as
\begin{eqnarray}
 \frac{|y_\nu y_\nu|^2}{ M_{H^+}^4}  + \frac{|y_\nu y_\nu|^2}{4}
   \left(\frac{1}{M_{H_{\nu}^0}^2} + \frac{1}{M^2_{A_{\nu}^0}} \right)^2 
   < \sqrt{\frac{g_*}{90}}\frac{64\pi^4 }{3\zeta(3) T_{QH}^3 M_P}
   = \frac{1}{(3.3 \, {\rm TeV})^4}\left(\frac{0.1 \,{\rm GeV}}{T_{QH}}\right)^3 .
\label{constraint:DR}
\end{eqnarray}
Therefore the neutrinophilic Higgs must be heavier than around $3\tev$
 for order of unity Yukawa couplings.
We note that the similar bound has already been obtained
 but $H_{\nu}^0$ and $A_{\nu}^0$ contributions were missing in Ref.~\cite{Davidson:2009ha,Mitropoulos:2013fla}. 
Here we have re-estimated and corrected it.

\section{dark matter}
\label{Sec:darkmatter}

The lightest RH sneutrino is stable when R-parity is preserved and
 can be a good candidate for dark matter.
The possibility in the neutrinophilic Higgs model was suggested in Ref.~\cite{Choi:2012ap,Choi:2013fva}
 by two of the present authors. 
In this section we generalise the previous results 
 considering the benchmark points in the previous section and examine the cosmological and astrophysical phenomenon.

\subsection{Relic density of dark matter}

Due to the large Yukawa coupling in \eq{Yukawa},
 the RH sneutrino interacts with fermions and Higgsinos efficiently so that they could be
 in the thermal equilibrium at high temperature. 
In the rapidly expanding early Universe,
 those RH sneutrinos decouple from the thermal plasma and the comoving abundance
 is conserved after that.
The relic density of WIMPs is determined by the annihilation cross section
 which determines the freeze-out temperature of DM. 
However, for complex fields, there might be the non-vanishing DM asymmetry.
With a large DM asymmetry, the final relic density of DM
 may depend on the annihilation cross section and
 the DM asymmetry~\cite{Barr:1990ca,Barr:1991qn,Kaplan:1991ah,Thomas:1995ze,Hooper:2004dc,Kitano:2004sv,Kaplan:2009ag}. 
This is the case for light RH sneutrino DM in our scenario. 

The annihilation cross section of RH sneutrino DM is dominantly determined
 by the annihilations into the leptons, that is given in partial wave expansion
 method by~\cite{Choi:2012ap,Choi:2013fva}
\begin{equation}
\langle \sigma v\rangle_{f\bar{f}} = \sum_f \left(
 \frac{y_\nu^4}{16\pi} \frac{m_f^2}{(M^2_{\tilde N} + M^2_{\tilde{H}_\nu} )^2}
 + \frac{ y_{\nu}^4}{8 \pi }
    \frac{M^2_{\tilde{N}}}{ ( M_{\tilde{N}}^2 + M_{\tilde{H_{\nu}}}^2 )^2 }\frac{T}{ M_{\tilde{N}} } + ... \right).
    \label{treelevel}
\end{equation}
There is another subdominant contribution from the induced annihilation into photons,   
\dis{
\VEV{\sigma v}_{2\gamma}
=\frac{|M|^2_{2\gamma} }{32\pi M^2_{\tilde N} }
 \simeq \frac{ \alpha^2_{\rm em}}{8\pi^3}  \frac{y_\nu^4 (A_\nu^2+\mu'^2)^2}{M_{\tilde l}^4} \frac{4}{M^2_{\tilde N}},\label{sig2g}
}
 where we used the approximation of $M_{H_\nu}=M_{H_\nu'}=M_{\tilde l}$ for simplicity
and  the soft term ${\cal L} = y_\nu A_{\nu} \tilde{L} H_\nu \tilde{N} + h.c. $.
In fact, it gives small subdominant contribution to determine the relics density of DM
 in our consideration with ${y_\nu^4 (A_\nu^2+\mu'^2)^2}\simeq M_{\tilde l}^4 $.

Since the RH sneutrinos were in the thermal equilibrium,
 the asymmetry could be generated from non-zero baryon asymmetry during the sphaleron process. 
The asymmetry of RH sneutrinos will depend on the specific model of baryogenesis, mass spectrum
 and the electroweak phase transition. 
In the simple case, the leptonic asymmetry is expected to be the order of baryon asymmetry,
 as $10^{-10}$~\cite{Mitropoulos:2013fla}. 
In this paper, in order to see the asymmetry dependence, 
 we treat it as a free parameter taking a value of a certain range.
 
For the WIMP with a non-vanishing asymmetry, the resulting relic density can be estimated
 by~\cite{Graesser:2011wi,Iminniyaz:2011yp}
\begin{eqnarray}
&& \Omega_{\rm DM} h^2 = 2.8 \times 10^8 \frac{m_{\rm DM}}{{\rm GeV}}
 (Y_{\rm DM}+Y_{\rm \overline{ DM}}) .
\end{eqnarray}
Here 
\dis{
 Y_{\rm DM}+Y_{\rm \overline {DM}}
 = \frac{C}{1-\exp[-C \lambda J[x_F] ]}
 + \frac{C}{\exp[C \lambda J[{\overline x_F}]]-1},
}
 with 
\begin{eqnarray}
\lambda &=& 4\pi\sqrt{\frac{g_*}{90}}M_P m_{\rm DM}, \\
J[x_F] &=& \int_{x_F}^\infty \langle\sigma v\rangle x^{-2}dx, \\
x &=& m_{\rm DM}/T,
\end{eqnarray}
where $x_F ({\overline x_F})$ denotes
 the value of $x$ at the ``freeze out'' time of (anti-)dark matter particle.
The asymmetry of dark matter is given by
\begin{equation}
 C = \frac{n_{\rm DM} -n_{\rm \overline {DM}}}{s}, \label{ADMC}.
\end{equation}
\begin{figure}[!t]
  \begin{center}
  \begin{tabular}{c}
   \includegraphics[width=0.7\textwidth]{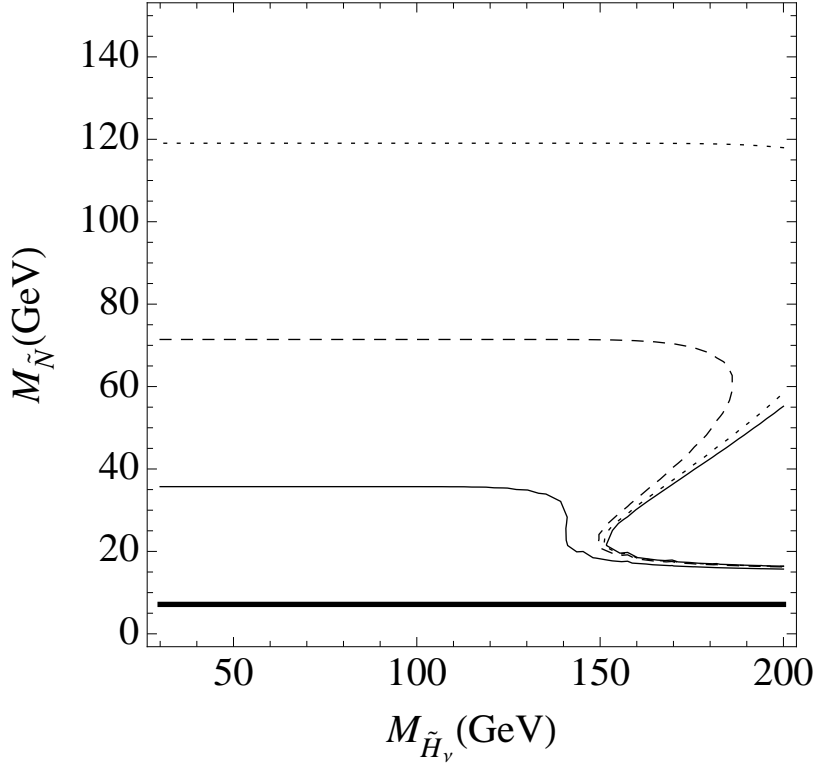}
  \end{tabular}
  \end{center}
  \caption{The contour plot of the DM asymmetry  $C=(5\times 10^{-11}, 10^{-11}, 5\times 10^{-12}, 3\times 10^{-12},10^{-12})$ with (thick solid, solid, dashed, dotted, solid lines) given in \eq{ADMC}, to give correct relic density for DM in the plane of its mass and Higgsino mass. }
  \label{fig:DMann}
\end{figure}

In the figure~\ref{fig:DMann} we show the contour plot of the corresponding DM asymmetry to give the correct relic density of DM. For a given asymmetries $C=(5\times 10^{-11}, 10^{-11}, 5\times 10^{-12}, 3\times 10^{-12},10^{-12})$,  the required amount of DM is obtained on the  corresponding (thick solid, solid, dashed, dotted, solid) lines in the plane of DM mass and Higgsino mass.

For a smaller Higgsino mass,
 the annihilation cross section is large enough to annihilate the DM-anti DM pairs and
 only the asymmetry remains.  
In this case, the abundance is determined by the asymmetry and the relic density is proportional to the DM mass and the asymmetry. 
We can see this for the Higgsino mass between $80\gev$ and $140\gev$,
 where the contour lines are parallel and horizontal,
 and the mass of dark matter is inversely proportional to the asymmetry.
For large Higgsino mass, the annihilation cross section is too small to annihilate all anti DM.
In this case, the relic density is determined by the annihilation cross section
 as the usual WIMP, being independent from the asymmetry. 
That can be seen, in the Figure,
 around Higgsino mass larger than $ 150\gev$ and RH sneutrino mass smaller
 than $60\gev$, where contour lines are not horizontal.

Compared to Fig.~\ref{fig:amu2} and ~\ref{fig:amu3},
 the region compatible to explain the muon anomalous magnetic moment corresponds
 to Higgsino mass between $40\gev$ and $120\gev$  
 and the DM relic density is dominantly determined by the DM asymmetry larger than around $C=5\times 10^{-12}$ . 

\subsection{Dark matter scattering cross section with a nucleon}

\begin{figure}[!t]
  \begin{center}
  \begin{tabular}{c}
   \includegraphics[width=0.7\textwidth]{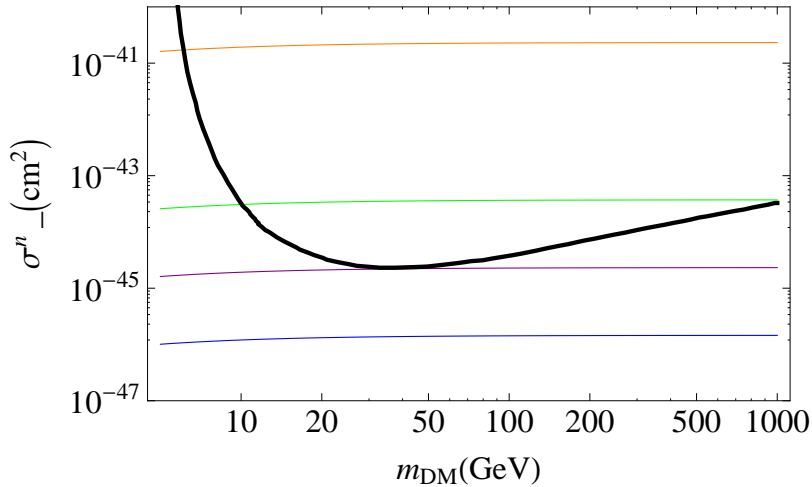}
  \end{tabular}
  \end{center}
  \caption{The scattering cross section between the lightest RH sneutrino and a neutron
 for $(S^T y_{\nu})A_{\nu}/M_{H_{\nu}} = 0.5, 1, 2, 10$ from the bottom to the top. The current LUX bound on it is also shown
 with the black thick line.}
  \label{fig:DMDD}
\end{figure}

An elastic scattering of RH sneutrino with nuclei may induce a signal
 in the direct detection experiments.
The most relevant process for our sneutrino DM is through the $Z$-boson exchange
 even this is not LH but RH~\cite{Choi:2012ap,Choi:2013fva},
 in contrast with most of other RH sneutrino DM models
 where the Higgs bosons exchange is dominant~\cite{Cerdeno,Choi:2012ba,Kang:2011wb,Kanemura:2014cka}. 
The DM-DM-$Z$ vertex is induced by one-loop digram involving neutrinophilic Higgs boson.
Because of the $Z$-boson exchange,
 DM mostly scatters with only neutrons rather than protons.
In Fig.~\ref{fig:DMDD}, we show
 the scattering cross section with a neutron for $(S^T y_{\nu})A_{\nu}/M_{H_{\nu}} = 0.1, 1, 2, 10$ from the bottom to the top, 
 and the current bounds by direct detection experiments, especially,
 the LUX experiment~\cite{Akerib:2013tjd} with the black thick line. The experimental bound shown in the Figure is re-estimated
 for our DM which scatters off with only neutrons.
Here $A_\nu$ is the trilinear soft coupling defined after \eq{sig2g}.
%


\subsection{Indirect signal from sneutrino dark matter}

For the asymmetric DM, their annihilation in the galaxy is negligible. 
However, they can scatter off the cosmic rays and produce secondary particles such as gamma-ray or neutrinos~\cite{Profumo:2011jt}. 
Since we are considering the large Yukawa couplings, the indirect signature might be very promising or even harmful. 
However, in our benchmark point 2 and 3,
 the Yukawa coupling between the lightest RH sneutrino ($N_2$) and
 the electron is quite negligible as a consequence of suppressing LFV and therefore the indirect signature is also  very suppressed.

For symmetric DM case with the heavy $\tilde{H}_{\nu}^{\pm}$,
 the DM annihilation signal can be seen most likely as a gamma-ray line
 because annihilation into a fermion pair is helicity suppressed~\cite{Choi:2012ap}.

\subsection{Decay of Cosmic Neutrino Background}

As the muon can decay to the electron through the charged neutrinophilic Higgs(ino), and sneutrino loops, 
 the heavier neutrino can decay to the lighter one through similar diagrams. 
Assuming that the mass of lighter one is much smaller than that of the heavier one, the decay rate of neutrino is
\dis{
\Gamma(\nu_j\rightarrow \nu_i + \gamma) =\frac{e^2}{4\pi}m_{\nu_j}^5 (|(A^\nu_2)_{ij}|^2+|(A^\nu_3)_{ij}|^2).
}
where 
\dis{
(A_2^\nu)_{i j} &\simeq
 \frac{1}{32\pi^2}  \left(  \sum_{\alpha}
 \frac{ (  y_\nu)_{i \alpha} (y_\nu)_{j\alpha}}{M_{\tilde{H}_\nu^-}^2}
 F\left(\frac{M_{\tilde{l}_{\alpha}}^2}{M_{\tilde{H}_\nu^-}^2}\right)
 - \sum_\beta \frac{ (y_{\nu})_{i \beta } (y_{\nu})_{j\beta} }{48 M_{H_\nu^-}^2}
  \right) ,  \quad\\
(A_3^\nu)_{i j} &= 0.
}
Although the GIM mechanism is not applied, so there is no suppression by the mass of charged leptons, 
the suppression by small neutrino mass as 5th powers is enough to satisfy the present constraint on the life-time of the neutrinos, which is
\dis{
\tau_\nu > 10^{12} \textrm{yrs},
} 
 from the analysis of the cosmic infrared background~\cite{Aboubrahim:2013gfa}.

\section{Conclusion}
\label{sec:con}

We have studied an extended supersymmetric model
 where neutrinos are Dirac particle and those masses are given
 by large neutrino Yukawa couplings and a small VEV of the neutrinophilic Higgs field.
Provided the lightest RH sneutrino is LSP as the dark matter candidate,
 we have examined various aspects of the model with Dirac Yukawa couplings
 of the order of unity.

By only considering the muon decay width, it turns out that
 the neutrinophilic Higgs bosons must be heavier than several hundreds GeV
 and some supersymmetric particles among RH sneutrinos
 and neutrinophilic Higgsino need to be heavier than several hundreds GeV.
In fact, we have found that the neutrinophilic Higgsino and
 one of the RH sneutrino can be relatively light of the order of $100\gev$
 if the other two RH neutrinos are heavy enough.
The current collider experiment, most importantly the LHC, constraints require a
 viable parameter space; $\tilde{H}_{\nu}^{\pm}$ is heavy enough,
 or the mass difference of $\tilde{H}_{\nu}^{\pm}$ and the lightest RH sneutrino 
 is smaller than about $50$ GeV, according to Ref.~\cite{Guo:2013asa}.
The LEP constrains
 the $\tilde{H}_{\nu}^{\pm}-\tilde{N}_{DM}-e$ coupling to be less than about $0.6$.

For general mixing of RH sneutrinos,
 due to the large flavor mixings of neutrino sector, the
 lepton flavor violating processes induced through neutrino Yukawa interactions
 are also typically as large as $10^{-6}$ in the decay branching ratio, 
 with new particles of ${\cal O}(100)$ GeV mass.
Only with appropriately tuned RH sneutrino mixings, we can avoid the LFVs.

With the chosen parameters, we found that the deviation of the muon $g-2$ can be explained
 with a relatively large lightest neutrino mass around $m_1\simeq 0.05\ev$ and the lightest RH sneutrino
  and  Higgsino mass, $M_{\tilde{N}}= 10 - 100$ GeV and $M_{\tilde{H}_{\nu}}=60 - 160$ GeV respectively.
In other words, if the muon $g-2$ is explained in this model,
 then $m_1$ can not be so small.
As a result of tuned RH sneutrino mixings, 
 the $\tilde{H}_{\nu}^{\pm}-\tilde{N}_{DM}-e$ coupling is almost vanishing,
 which means the LEP data does not significantly constrain this model and
 the international linear collider also would not be able to produce a mono-photon signal,
 while the $\tilde{H}_{\nu}^{\pm}-\tilde{N}_{DM}-\mu$ coupling is about unity.
A muon collider can easily test this model
 if it will be indeed constructed, as the Fermilab plans~\cite{MAP}.

In this muon $g-2$ favored parameter region,
 the DM relic density is explained by RH sneutrino  with
 the asymmetry of $C \sim 5\times 10^{-12}$.
If we do not mind the deviation of the muon $g-2$,
 RH sneutrino dark matter could be an usual WIMP
 with heavier $\tilde{H}_{\nu}^{\pm}$.

\section*{Acknowledgments}

K.-Y.C. appreciates Asia Pacific Center for Theoretical Physics for the support to the Topical Research Program.
K.-Y.C. was supported by the Basic Science Research Program through the National Research Foundation of Korea (NRF) funded by the Ministry of Education, Science and Technology Grant No. 2011-0011083. 
O.S. was supported in part by the Grant-in-Aid for Scientific Research 
 on Innovative Areas No.~26105514 from
 the Ministry of Education, Culture, Sports, Science and Technology in Japan.
C.S.S. is supported in part by DOE grants doe-sc0010008, DOE-ARRA- SC0003883, and DOE-DE-SC0007897.

\section*{Appendix}

\subsection{Formula for the muon $g-2$}

For the Lagrangian
\dis{
\mathcal{L} = \bar{L} (\lambda_v +i\lambda_a \gamma_5) f \, S +h.c.,
}
where $f$ is charged fermion, the anomalous magnetic dipole moment of the charged lepton $L$ with charge $Q$ is given by~\cite{Hisano:1995cp,Hisano:2001qz}
\dis{
\delta a_L=\frac{m_L}{8\pi^2} \int_0^1 dz \frac{( |\lambda_v|^2  +|\lambda_a|^2) z(1-z)^2 m_L + ( |\lambda_v|^2  -|\lambda_a|^2) (1-z)^2 m_f}{-z(1-z) m_L^2 +(1-z)m_f^2 +z m_s^2}.
}
For neutrinophilic Higgs model with Lagrangian
\dis{
\mathcal{L} = y \tilde{N} \bar{L} P_R \tilde{H_\nu} +h.c.
} 
we obtain
\begin{eqnarray}
\delta a_L &=& \frac{|y|^2m_L^2}{16\pi^2}\int_0^1 dz \frac{ z(1-z)^2 }{-z(1-z) m_L^2 +(1-z)m_f^2 +z m_s^2} \nonumber \\
&\simeq & \frac{|y|^2m_L^2}{16\pi^2 m_f^2} \int _0^1 dz  \frac{z(1-z)^2}{1-z + z (m_s^2/m_f^2)}.
\end{eqnarray}

\subsection{Auxiliary functions}

\begin{eqnarray}
F(x) &=& \int_0^1 \frac{z(1-z)^2}{(1-z)+zx} \nonumber\\
 &=& \frac{1-6x+3x^2+2x^3-6x^2\ln(x)}{6(x-1)^4},
\end{eqnarray}
\begin{eqnarray}
 F_2(x,y) = \frac{1}{x^2-y^2}\ln\left(\frac{x^2}{y^2}\right) ,
\end{eqnarray}
\begin{eqnarray}
 &&F_3(x,y,z,w) \nonumber\\
&& = \frac{ x^4(z^2-w^2)\log(-x^2)+w^4(x^2-z^2)\log(-w^2)+z^4(w^2-x^2)\log(-z^2)}{(x^2-y^2)(z^2-w^2)(x^2-z^2)(x^2-w^2) } \nonumber\\
& &\quad
 - \frac{ y^4(z^2-w^2)\log(-y^2)+ w^4(y^2-z^2)\log(-w^2)+z^4(w^2-y^2)\log(-z^2)}{(x^2-y^2)(z^2-w^2)(z^2-y^2)(w^2-y^2) } .
\end{eqnarray}
%




\begin{thebibliography}{99}

\bibitem{Aad:2012tfa}
  G.~Aad {\it et al.}  [ATLAS Collaboration],
  Phys.\ Lett.\ B {\bf 716} (2012) 1
  [arXiv:1207.7214 [hep-ex]].

\bibitem{Chatrchyan:2012ufa}
  S.~Chatrchyan {\it et al.}  [CMS Collaboration],
  Phys.\ Lett.\ B {\bf 716} (2012) 30
  [arXiv:1207.7235 [hep-ex]].

\bibitem{Seesaw}
T.~Yanagida,
 in \textit{Proceedings of Workshop on the Unified Theory and
 the Baryon Number in the Universe}, Tsukuba, Japan,
 edited by A.~Sawada and A.~Sugamoto (KEK, Tsukuba, 1979), p 95;\\
M.~Gell-Mann, P.~Ramond, and R.~Slansky,
 in \textit{Supergravity},
 Proceedings of Workshop, Stony Brook, New York, 1979, edited by
 P.~Van~Nieuwenhuizen and D.~Z.~Freedman
 (North-Holland, Amsterdam, 1979), p 315;\\
R.~N.~Mohapatra and G.~Senjanovic, Phys. Rev. Lett. {\bf 44}, 912 (1980).

\bibitem{Ma}
  E.~Ma, Phys. Rev. Lett. {\bf 86}, 2502 (2001). 

\bibitem{Wang}
  F.~Wang, W.~Wang and J.~M.~Yang, Europhys.\ Lett.\ {\bf 76}, 388 (2006).

\bibitem{Nandi}
  S.~Gabriel and S.~Nandi, Phys. Lett. B {\bf 655}, 141 (2007).

\bibitem{Choi:2012ap}
  K.~-Y.~Choi and O.~Seto,
  Phys.\ Rev.\ D {\bf 86} 043515 (2012)
   [Erratum-ibid.\ D {\bf 86} 089904 (2012)].
  [arXiv:1205.3276 [hep-ph]].
\bibitem{Choi:2013fva}
  K.~-Y.~Choi and O.~Seto,
  Phys.\ Rev.\ D {\bf 88} (2013) 035005
  [arXiv:1305.4322 [hep-ph]].

\bibitem{Mitropoulos:2013fla}
  P.~Mitropoulos,
  JCAP {\bf 1311} (2013) 008
  [arXiv:1307.2823].

\bibitem{HabaSeto}
  N.~Haba and O.~Seto,
  Prog.\ Theor.\ Phys.\ {\bf 125}, 1155 (2011); Phys.\ Rev.\ D {\bf 84}, 103524 (2011).
%
\bibitem{HSY}
  N.~Haba, O.~Seto and Y.~Yamaguchi,
  Phys.\ Rev.\ D {\bf 87} (2013) 123540
  [arXiv:1305.2484 [hep-ph]].
%

\bibitem{PDG} 
  J.~Beringer {\it et al.}  [Particle Data Group Collaboration],
  Phys.\ Rev.\ D {\bf 86}, 010001 (2012).


\bibitem{Ade:2013zuv} 
  P.~A.~R.~Ade {\it et al.}  [Planck Collaboration],
  arXiv:1303.5076 [astro-ph.CO].

\bibitem{0809.5221} 
  T.~Fukuyama and K.~Tsumura,
  arXiv:0809.5221 [hep-ph].
%
\bibitem{Fet} 
  W.~Fetscher, H.-J.~Gerber, and K.~F.~Johnson, Phys.\ Lett.\ B {\bf 173}, 102 (1986).
%
\bibitem{Herczeg:1997bu} 
  P.~Herczeg,
  Los Alamos Sci.\  {\bf 25}, 128 (1997).
%

\bibitem{Marciano:1999ih} 
  W.~J.~Marciano,
  Phys.\ Rev.\ D {\bf 60}, 093006 (1999).
%
\bibitem{Erler:2004cx} 
  J.~Erler and M.~J.~Ramsey-Musolf,
  Prog.\ Part.\ Nucl.\ Phys.\  {\bf 54}, 351 (2005).
 [hep-ph/0404291].
%

\bibitem{charm} 
  P.~Vilain {\it et al.}  [CHARM-II Collaboration],
  Phys.\ Lett.\ B {\bf 364}, 121 (1995).

\bibitem{Davidson:2009ha}
  S.~M.~Davidson and H.~E.~Logan,
  Phys.\ Rev.\ D {\bf 80}, 095008 (2009); 
  Phys.\ Rev.\ D {\bf 82}, 115031 (2010).
\bibitem{Haba:2011nb}
  N.~Haba and K.~Tsumura,
  JHEP {\bf 1106}, 068 (2011).

\bibitem{LEPSUSY}
LEP SUSY Working Group (ALEPH, DELPHI, L3, OPAL), Notes LEPSUSYWG/01-03.1, 04-01.1, http://lepsusy.web.cern.ch/lepsusy/Welcome.html.
%
\bibitem{Fox:2011fx} 
  P.~J.~Fox, R.~Harnik, J.~Kopp and Y.~Tsai,
  Phys.\ Rev.\ D {\bf 84}, 014028 (2011).

\bibitem{Guo:2013asa} 
  J.~Guo, Z.~Kang, J.~Li, T.~Li and Y.~Liu,
  arXiv:1312.2821 [hep-ph].



\bibitem{Adam:2013mnn}
  J.~Adam {\it et al.}  [MEG Collaboration],
  Phys.\ Rev.\ Lett.\  {\bf 110} (2013) 20,  201801
  [arXiv:1303.0754 [hep-ex]].
%
\bibitem{Aubert:2009ag}
  B.~Aubert {\it et al.}  [BaBar Collaboration],
  Phys.\ Rev.\ Lett.\  {\bf 104} (2010) 021802
  [arXiv:0908.2381 [hep-ex]].

\bibitem{Bennett:2006fi}
  G.~W.~Bennett {\it et al.}  [Muon G-2 Collaboration],
  Phys.\ Rev.\ D {\bf 73} (2006) 072003
  [hep-ex/0602035].
\bibitem{Hagiwara:2011af}
  K.~Hagiwara, R.~Liao, A.~D.~Martin, D.~Nomura and T.~Teubner,
  J.\ Phys.\ G {\bf 38} (2011) 085003
  [arXiv:1105.3149 [hep-ph]].


\bibitem{Barr:1990ca}
  S.~M.~Barr, R.~S.~Chivukula and E.~Farhi,
  Phys.\ Lett.\  B {\bf 241}, 387 (1990).
\bibitem{Barr:1991qn}
  S.~M.~Barr,
  Phys.\ Rev.\  D {\bf 44}, 3062 (1991).
\bibitem{Kaplan:1991ah}
  D.~B.~Kaplan,
  Phys.\ Rev.\ Lett.\  {\bf 68}, 741 (1992).
\bibitem{Thomas:1995ze}
  S.~D.~Thomas,
  Phys.\ Lett.\  B {\bf 356}, 256 (1995).
\bibitem{Hooper:2004dc}
  D.~Hooper, J.~March-Russell and S.~M.~West,
  Phys.\ Lett.\  B {\bf 605}, 228 (2005).
\bibitem{Kitano:2004sv}
  R.~Kitano and I.~Low,
  Phys.\ Rev.\  D {\bf 71}, 023510 (2005).
\bibitem{Kaplan:2009ag}
  D.~E.~Kaplan, M.~A.~Luty and K.~M.~Zurek,
  Phys.\ Rev.\  D {\bf 79}, 115016 (2009).

\bibitem{Graesser:2011wi}
  M.~L.~Graesser, I.~M.~Shoemaker and L.~Vecchi,
  JHEP {\bf 1110} (2011) 110
  [arXiv:1103.2771 [hep-ph]].
  
\bibitem{Iminniyaz:2011yp}
  H.~Iminniyaz, M.~Drees and X.~Chen,
  JCAP {\bf 1107} 003 (2011).

\bibitem{Cerdeno}
  D.~G.~Cerdeno, C.~Munoz and O.~Seto,
  Phys.\ Rev.\ D {\bf 79}, 023510 (2009);\\
  D.~G.~Cerdeno and O.~Seto,
  JCAP {\bf 0908} (2009) 032
  [arXiv:0903.4677 [hep-ph]];\\
  D.~G.~Cerdeno, J.~-H.~Huh, M.~Peiro and O.~Seto,
  JCAP {\bf 1111}, 027 (2011);\\
  D.~G.~Cerdeno, V.~Martín-Lozano and O.~Seto,
  JHEP {\bf 1405} (2014) 035
  [arXiv:1311.7260 [hep-ph]];\\
  D.~G.~Cerdeno, M.~Peiro and S.~Robles,
  arXiv:1404.2572 [hep-ph].
\bibitem{Choi:2012ba}
  K.~-Y.~Choi, E.~J.~Chun and C.~S.~Shin,
  Phys.\ Lett.\ B {\bf 723} (2013) 90
  [arXiv:1211.5409 [hep-ph]].
\bibitem{Kang:2011wb}
  Z.~Kang, J.~Li, T.~Li, T.~Liu and J.~Yang,
  arXiv:1102.5644 [hep-ph]; \\
  J.~Guo, Z.~Kang, T.~Li and Y.~Liu,
  JHEP {\bf 1402} (2014) 080
  [arXiv:1311.3497 [hep-ph]].
\bibitem{Kanemura:2014cka}
  S.~Kanemura, N.~Machida and T.~Shindou,
  arXiv:1405.5834 [hep-ph].

\bibitem{Akerib:2013tjd}
  D.~S.~Akerib {\it et al.}  [LUX Collaboration],
  Phys.\ Rev.\ Lett.\  {\bf 112} (2014) 091303
  [arXiv:1310.8214 [astro-ph.CO]].

\bibitem{Profumo:2011jt}
  S.~Profumo and L.~Ubaldi,
  JCAP {\bf 1108} (2011) 020
  [arXiv:1106.4568 [hep-ph]].
%
\bibitem{Aboubrahim:2013gfa}
  A.~Aboubrahim, T.~Ibrahim and P.~Nath,
  Phys.\ Rev.\ D {\bf 88} (2013) 013019
  [arXiv:1306.2275 [hep-ph]].

\bibitem{MAP}
http://map.fnal.gov/

\bibitem{Hisano:1995cp}
  J.~Hisano, T.~Moroi, K.~Tobe and M.~Yamaguchi,
Phys.\ Rev.\ D {\bf 53} (1996) 2442  [hep-ph/9510309].  
\bibitem{Hisano:2001qz} 
  J.~Hisano and K.~Tobe,
Phys.\ Lett.\ B {\bf 510}, 197 (2001)  [hep-ph/0102315]. 
\end{thebibliography}
\end{document}